\documentclass{aa}  
\usepackage{graphicx}
\usepackage{txfonts}

\usepackage[breaklinks=true, colorlinks=true, citecolor=blue, linkcolor=blue,urlcolor=blue]{hyperref}
\begin{document} 

   \title{Globular clusters in M\,104: Tracing kinematics and metallicities from the centre to the halo}

   \subtitle{}
   \titlerunning{M\,104 globular clusters from the centre to the halo}

   \author{Katja Fahrion\inst{1}
            \and
            Michael A. Beasley\inst{2}\fnmsep\inst{3}\fnmsep\inst{4}
            \and
            Eric Emsellem\inst{5}
            \and
            Anastasia Gvozdenko\inst{6}
            \and
            Oliver M\"uller\inst{7}\fnmsep\inst{8}\fnmsep\inst{9}
            \and
            Marina Rejkuba\inst{5}
            \and
            Ana L. Chies-Santos\inst{10}
   }

    \institute{        
            Department of Astrophysics, University of Vienna, T\"{u}rkenschanzstra{\ss}e 17, 1180 Wien, Austria.\\
            \email{katja.fahrion@univie.ac.at}
        \and 
            Instituto de Astrofísica de Canarias, Calle V\'{i}a L\'{a}ctea, E-38206 La Laguna, Spain.          
        \and  
            Departamento de Astrof\'{i}sica, Universidad de La Laguna, E-38206 La Laguna, Spain.
        \and  
            Centre for Astrophysics and Supercomputing, Swinburne University, John Street, Hawthorn VIC 3122, Australia.
        \and 
            European Southern Observatory, Karl-Schwarzschild-Stra{\ss}e 2, 85748 Garching bei M\"unchen, Germany
        \and 
        Department of Physics, Centre for Extragalactic Astronomy, Durham University, South Road, Durham DH1 3LE, UK
        \and 
         Institute of Physics, Laboratory of Astrophysics, Ecole Polytechnique Fédérale de Lausanne (EPFL), 1290 Sauverny, Switzerland             
           \and 
           Institute of Astronomy, Madingley Rd, Cambridge CB3 0HA, UK
        \and 
        Visiting Fellow, Clare Hall, University of Cambridge, Cambridge, UK
        \and 
        Instituto de F\'{i}sica, Universidade Federal do Rio Grande do Sul (UFRGS), Av. Bento Gon\c{c}alves, 9500, Porto Alegre, RS 91501-970, Brazil
             }
    
   \date{\today}
 
  \abstract  
  {As ancient star clusters, globular clusters (GCs) are regarded as powerful tracers of 
galaxy evolution and assembly. Due to their brightness and compact sizes, GCs are employed to probe the kinematics and stellar population properties of galaxies, from the central regions out into the halo where the underlying stellar light becomes too faint for spectroscopic studies.
In this work, we present a comprehensive study of the GC system of M\,104 (NGC\,4594, also known as the Sombrero galaxy) based on literature spectroscopic catalogues and newly collected data from Very Large Telescope (VLT) MUSE integral-field spectroscopy combined with multi-object spectroscopy from VLT FLAMES and OSIRIS at the Gran Telescopio de Canarias (GTC). 
We present a new catalogue of 499 GCs with radial velocity measurements that span from the inner disc region out to $\sim$ 70 kpc (24\arcmin). In addition to velocities, we measured metallicities from the MUSE, OSIRIS, and FLAMES spectra of 190 GCs. Together with literature values, we collected a sample of 278 metallicities. We found good agreement between the velocity and metallicity measurements of GCs observed with multiple instruments. Studying GC kinematics with a simple model confirms a decreasing velocity dispersion profile and low rotation velocities. The blue GCs appear to be more dispersion-dominated, while the red GCs follow the kinematics of the stars more closely.
We find a large scatter of GC metallicities with distance from the centre, and metal-rich GCs are found over all radii. We discuss how the GC metallicity distribution with a broad metal-poor component likely reflects the complex assembly history of M\,104.}

   \keywords{Galaxies: star clusters: general -- galaxies: individual: M104}
               
   \maketitle

\section{Introduction}

As ancient star clusters with typical ages well above 10 Gyr, globular clusters (GCs) have long been regarded as powerful tracers of galaxy formation and assembly (see \citealt{Forbes2018} or \citealt{Beasley2020} for reviews). With their masses on the higher end of the stellar cluster population ($10^4$ - $10^6 M_\sun$) and compact sizes ($R_\text{eff} \sim 3$ pc), GCs are observable in the halo regions of distant galaxies. High-quality photometric GC catalogues from space-based imaging or multi-band ground-based observations are available for hundreds of galaxies (e.g. \citealt{LeCooper2025, Lim2025}), providing luminosities, sizes, and colours for thousands of GCs (see \citealt{Jordan2007, Harris2013, Harris2017}). Complementary spectroscopy is required to use GCs as tracers of galaxy properties and mass assembly since it enables measurement of line-of-sight (LOS) velocities and stellar population properties such as metallicities or light element abundances (e.g. \citealt{Pota2013, Forbes2017, Chaturvedi2022}). 

In this paper, we revisit the GC system of M\,104 (NGC\,4594), also known as the Sombrero galaxy. At a distance of $D = 9.55 \pm 0.13 \pm 0.31$ Mpc \citep{McQuinn2016}, M\,104 has been studied extensively due to its peculiar morphology with a nearly edge-on orientation and prominent dust lane \citep{Emsellem1995, Sutter2022}. M\,104 is traditionally classified as an Sa galaxy; however given its prominent spheroidal component \citep{GadottiSanchezJanssen2012}, metal-rich halo \citep{Cohen2020}, and high specific frequency of GCs ($S_N \sim 2$; \citealt{RhodeZepf2004, Kang2022}), there is growing evidence that M\,104 is a lenticular early-type galaxy (see \citealt{Kang2022}). With a stellar mass of $M_\ast \sim 2 \times 10^{11} M_\sun$ \citep{Jardel2011, Karachentsev2020}, it is the brightest and most massive member of its group. Based on radial velocities of dwarf galaxies, the total mass of the group is $M_\text{tot} = (1.24 \pm 0.65) \times 10^{13} M_\sun$ \citep{Crosby2025}. 

The rich GC system of M\,104 has been extensively studied for many decades using both imaging and spectroscopy (e.g. \citealt{Wakamatsu1977, HarrisHarrisHarris1984, Bridges1992, Bridges1997, Forbes1997, Bridges2007}). The total number of GCs is estimated to be around 1500 - 2000 \citep{HarrisHarrisHarris1984, RhodeZepf2004, Kang2022}, which is about an order of magnitude more than the number of GCs around the Milky Way or M\,31. The GC colour distribution of M\,104 was found to be bimodal, with a large fraction of red metal-rich GCs \citep{Larsen2001, Larsen2002, AlvesBrito2011}. In 2003, M\,104 was observed with multiple pointings of the Advanced Camera for Surveys (ACS) on board the \textit{Hubble} Space Telescope (HST), enabling a 600\arcsec $\times$ 400\arcsec mosaic to be constructed. In these data, 659 GCs were identified \citep{Spitler2006}, and \cite{Harris2010} presented updated photometry and size measurements for 652 GCs in the inner region of M\,104. More recently, \cite{Kang2022} used ground-based imaging with the MegaCam instrument at the Canada-France-Hawaii Telescope and presented a catalogue of $\sim 2900$ candidates out to 35\arcmin\,($\sim$ 100 kpc). They estimated a total number of GCs of 1610 $\pm$ 30 based on the number of candidates above the turn-over magnitude. 

The most recent spectroscopic catalogue of GCs around the Sombrero galaxy was compiled by \cite{Dowell2014}. They combined their own observations of 51 GCs with the sample of 108 GCs from \cite{Bridges2007} and the catalogue of 259 sources from \cite{AlvesBrito2011}. The catalogue of \cite{Dowell2014} lists 579 unique sources, about 360 of them are confirmed GCs of M\,104 based on their radial velocities.

In this paper, we aim to build on the above mentioned works to present the largest catalogue of spectroscopically confirmed GCs around the Sombrero galaxy to date. Combining the literature compilation from \cite{Dowell2014} with new observations from the Multi Unit Spectroscopic Explorer (MUSE; \citealt{Bacon2014}) instrument at the Very Large Telescope (VLT), the Fibre Large Array Multi Element Spectrograph (FLAMES; \citealt{Pasquini2002}) instrument at the VLT, and the Optical System for Imaging and low-Intermediate-Resolution Integrated Spectroscopy (OSIRIS; \citealt{Cepa1998}) instrument at the Gran Telescopio Canarias (GTC), we report on a catalogue of 499 GCs. Section \ref{sect:data} describes those datasets, and we present our analysis methods in Sect. \ref{sect:methods}. Results are presented and discussed in Sect. \ref{sect:results}. We conclude in Sect. \ref{sect:conclusions}.

\section{Data}
\label{sect:data}
\begin{figure}
    \centering
    \includegraphics[width=0.4\textwidth]{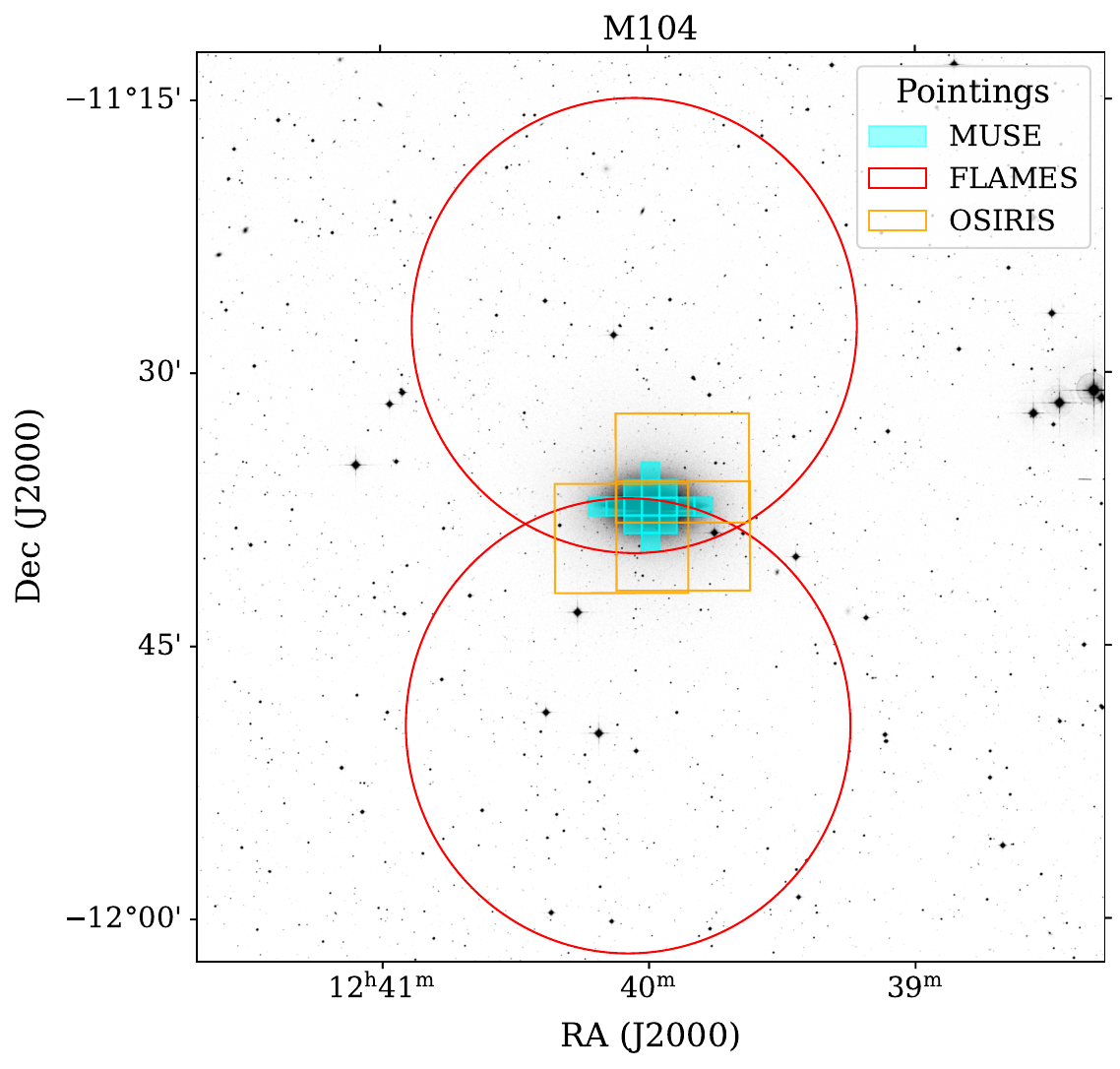}
    \caption{Footprints of the different datasets on a greyscale DSS2 image of M\,104. MUSE pointings are shown in cyan squares, each with a field of view of $1\arcmin \times 1\arcmin$. The red circles indicate the field of view of the FLAMES GIRAFFE instrument, with a diameter of 25\arcmin, and the orange rectangles show the field of view of OSIRIS, which spans $7.5\arcmin \times 6\arcmin$. The full greyscale image is 40\arcmin $\times$ 40\arcmin\,(110 kpc $\times$ 110 kpc).}
    \label{fig:pointings}
\end{figure}

In this work, we collected a new large sample of spectroscopically confirmed GCs using MUSE integral field spectroscopic data, FLAMES high resolution multi-object spectroscopy, and deep OSIRIS multi-object spectroscopy of bright GCs. This section describes the data reduction methods and Fig. \ref{fig:pointings} shows the observation footprints.

\subsection{MUSE data}
M\,104 was observed with MUSE as part of the science verification run (programme ID 60.A-9303, PI: Emsellem) in 2014 after MUSE was commissioned at the VLT. The MUSE data consist of 15 unique pointing positions covered by two exposures each along the disc as well as above and below around the central part of M\,104. The exposure time varies between 210 and 600 seconds per exposure, with shorter exposure times on the bright galaxy centre. The total exposure times increase outwards with the central exposure cumulating 420 seconds, while the two surroundings disc fields have 600 seconds. The next set of 8 fields (in the disc and off-plane) have 900 seconds, while the 4 most outer pointings have 1200 seconds. The average seeing was 1.2\arcsec, with some exposures reaching 0.45\arcsec and others $>$2.0\arcsec.

All data were reduced following a standard process and making use of python scripts wrapping around \textsc{esorex} commands and the Data Reduction Software developed by the MUSE consortium \citep{Weilbacher2020}.
The relative alignment of individual exposure was performed using the reference HST $F625W$ mosaic. Individual dedicated sky pointings were used to remove the background contribution after renormalisation of the derived sky spectrum. The final MUSE mosaic was assembled using the \textsc{esorex muse\_exp\_combine} command. We only considered the bluer part of the data cube up to 7400\,\AA\,due to stronger sky residuals at longer wavelengths. The spectra are sampled at 1.25\,\AA\,pix$^{-1}$ and with an instrumental resolution of $\sim$ 2.5\,\AA.

To detect the spectra of GCs in the MUSE data, we followed the methods described in \cite{Fahrion2020b}. As first step, we created a white light image using the median of 1000 wavelength slices from the MUSE data cube. Then, the bright galaxy background was removed using a median filter based on \textsc{photutils}' background subtraction functions. We chose a box size of 5 $\times$ 5 pixels with a filtersize of 3 $\times$ 3 pixels. In the background subtracted image, all sources 2 $\sigma$ above the background rms were detected using \textsc{photutils} \textsc{segmentation} detection routines with a FWHM of 4 pixels. To remove resolved background galaxies, only sources with eccentricity $< 0.7$ were kept, yielding 488 initial sources.

In the second step, we extracted the spectra of those sources using a Gaussian-weighted aperture with a FWHM of 4 pixels (0.8\arcsec). We also tested other values, but found that the influence on the final spectrum is minimal. For each source, a background spectrum was extracted using an annulus with inner radius of 8 (1.6\arcsec) and outer radius of 13 pixels (2.6\arcsec). Subtracting the background spectrum, we obtained 258 GC candidates with signal-to-noise ($S/N$) $>$ 2.5 pix$^{-1}$, as derived from the \textsc{estimateSNR} function of \textsc{PyAstronomy} \citep{pya} in a wavelength region between 6400 and 6500 \AA. Of those, 139 sources could be matched to the photometric GC catalogue of \cite{Harris2010} based on HST data. 
In the last step, the spectra were fitted with \textsc{pPXF} as described in Sect. \ref{sect:fitting_MUSE_spectra} to confirm membership to M\,104.

\subsection{FLAMES data}
The extensive GC system of M\,104 was observed with FLAMES GIRAFFE in MEDUSA mode (multi-object spectroscopy, programme ID 114.274W, PI: Fahrion) in January and February 2025 in two pointings, one to the north of the disc and one to the south. Each pointing has a diameter of 25\arcmin\,and the FLAMES fibers were placed on 221 GC candidates derived from the HST catalogues \citep{Spitler2006, Harris2010} and the catalogue from \cite{Kang2022} based on ground-based imaging. Additionally,  12 fibers were placed on empty sky regions. The HR21 (875nm) grating was used, which provides coverage of the calcium triplet region from 8483 to 8900 \AA\,at a resolution of R=18000 with a sampling of 0.05 \AA\,pix$^{-1}$. Each pointing was covered in ten exposures of 1365 seconds each, distributed over five observing blocks. 

The FLAMES data were reduced using a custom wrapper around the data reduction scripts of the FLAMES \textsc{esorex} pipeline\footnote{\url{https://www.eso.org/sci/software/pipelines/giraffe/giraffe-pipe-recipes.html}}. We followed mainly the standard processing steps according to the manual, including bias and flat fielding corrections, followed by a wavelength calibration. As recommended in the manual, we ran the wavelength calibration twice to obtain a custom slit geometry table and an improved wavelength solution. Afterwards, the science reduction is completed for each exposure. We chose the extraction method \textsc{`OPTIMAL'}, which gives an improved spectral extraction \citep{Horne1986}.

In a second step of the data reduction, the sky was subtracted. In the \textsc{esoreflex} pipeline, a master sky is created from all sky fibers and then subtracted. For bright sources, this approach is sufficient, but we chose to implement an improved approach to further reduce residuals for also fainter objects. Individual sky fibers were used to create a library of sky templates for each exposure. Then, we used the penalized pixel fitting method (\textsc{pPXF}; \citealt{Cappellari2004, Cappellari2017, Cappellari2023}), a full spectrum fitting method to fit spectra with user-supplied models, to fit each spectrum with a stellar model from the PHOENIX library \citep{Husser2013} and the sky templates as sky input. The PHOENIX models give synthetic stellar atmosphere spectra, provided at a spectral resolution of R=500'000, which we brought to the FLAMES resolution by convolution with a Gaussian. The full library covers a large range of stellar parameters, but we selected red giant stars with effective temperature between 4500 and 5000 K, surface gravity between log(g)= 2.0 and 4.0, and a range of metallicities between $-$1.5 and 0 dex. We note here that we only used the PHOENIX stellar spectra for the sky subtraction and did not consider the fits further. The best-fitting combination of sky templates are then subtracted and the spectra are saved. Compared with the master sky subtraction approach, the fitting with \textsc{pPXF} allows us to accurately scale the sky spectra and further reduce the sky line residuals. In addition to the sky subtraction, we also used the sky observations to measure the spectral resolution. For this, we fitted the strong sky lines in each sky fiber of each exposure, testing if the spectral resolution is stable across exposure and fibers. As also described in \cite{Beasley2025}, we did not find a trend with wavelength or exposure. The median FWHM of the sky lines is 0.54 $\pm$ 0.04 \AA\ across the different exposures. 

In the final step of the data reduction, the exposures for each individual source were combined, accounting for the barycentric velocity of each exposure and resampling all spectra on a common wavelength array. To get an improved combination, we used a median, weighted by the $S/N$ in each exposure. We also kept a version of the spectra without sky subtraction and used those to mask regions of strong sky contamination in the fits (see. Sect. \ref{sect:FLAMES_spec})

\subsection{GTC data}
75 GCs of M\,104 were observed with the GTC using the OSIRIS instrument in multi-object spectroscopy mode (programme IDs GTC24-19A, GTC89-20A, PI: Beasley). Three  fields were covered (Field 2: RA: 12:39:52.31, Dec.: $-$11:35:15.7; Field 3 : RA: 12:40:06.02, Dec.: $-$11:39:07.5; Field 4: RA: 12:39:52.16, Dec.: $-$11:38:59.8; J2000), using exposure times of 3 to 3.75 hours in 900 s individual exposures. Targets were selected from the HST ACS catalogue of \cite{Spitler2006} augmented with candidates from \cite{RhodeZepf2004}. The R1000B grating was used with a 0.6\arcsec\,slit,  giving a wavelength range of 3700--7500 \AA \,and spectral resolution with a FWHM of $\sim5$ \AA. 

Data were reduced using the GTC OSIRIS spectrograph package implemented in \textsc{PypeIt}\footnote{\url{https://github.com/pypeit/PypeIt}} \citep{PypeIt2020, pypeit:zenodo}. Standard data reduction procedures were followed, and the individual spectra were wavelength calibrated (in vacuum wavelengths), optimally extracted, corrected to the barycentric frame, and combined. The one dimensional spectra were flux- and extinction calibrated based on long-slit observations of the standard star HD289002 (Hiltner 600) observed in the same setup but with a 1.0\arcsec\,slit. We note that the flux calibration applied is only approximate -- mainly to correct the continuum shape -- due to the problem of precise calibrations of multi-slit spectra. The final $S/N$ ratio of the spectra are high, with median $S/N$ = 74 pix$^{-1}$.

\section{Spectroscopic analysis}
\label{sect:methods}
\begin{figure}
    \centering
    \includegraphics[width=0.49\textwidth]{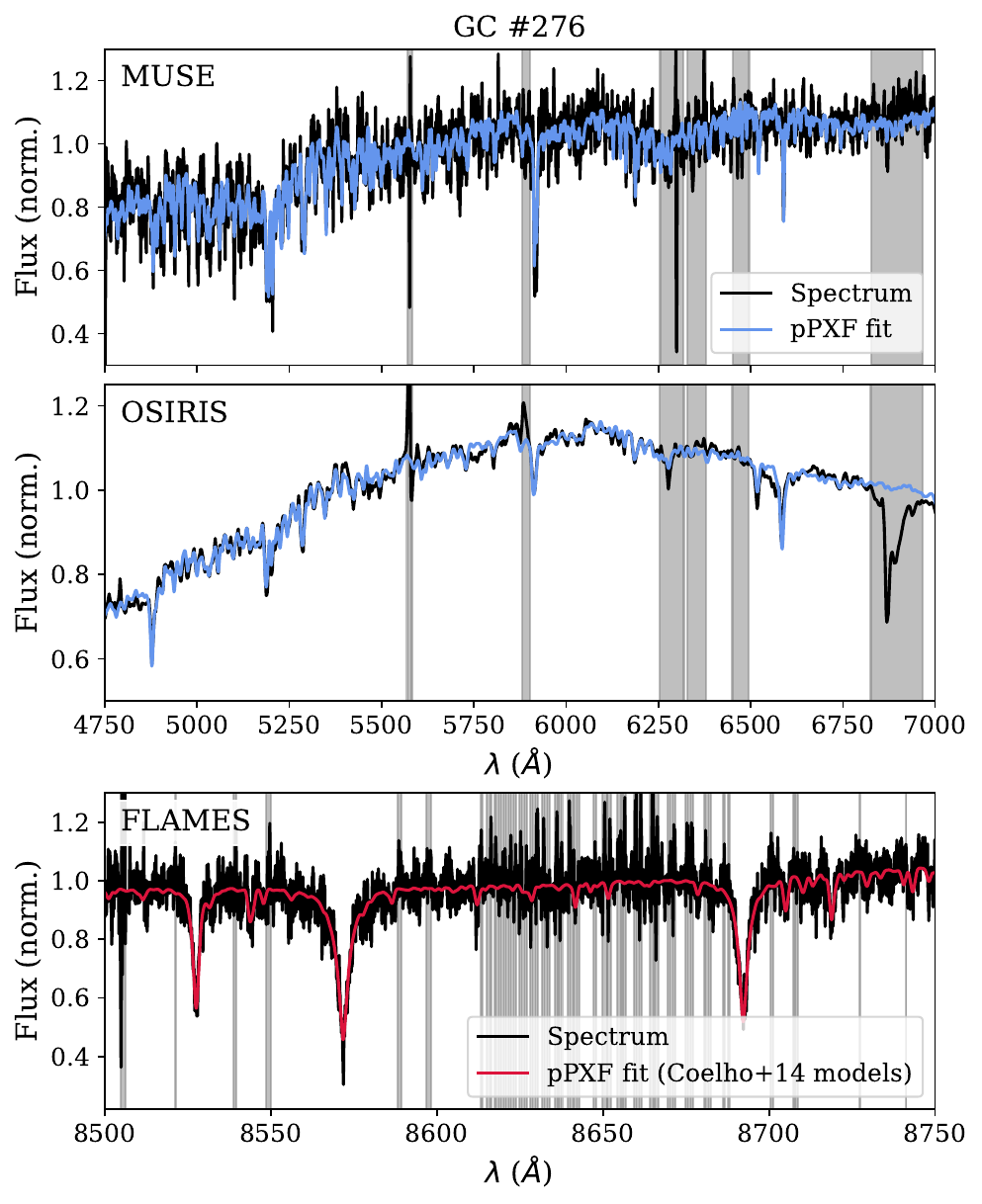}
    \caption{Spectra of GC \#276 for which spectra from MUSE (top), OSIRIS (middle), and FLAMES (bottom) are available. The black lines show the data, and the blue and red lines are the fits with \textsc{pPXF}. Shaded regions were excluded due to sky line residuals. The MUSE and OSIRIS spectra were fitted with XSL SSPs, while we used high resolution stellar models from \cite{Coelho2014} to fit the FLAMES spectrum shown here.}
    \label{fig:ACS15_Spectra}
\end{figure}

We fit all GC spectra with \textsc{pPXF} and tested different models. Figure \ref{fig:ACS15_Spectra} shows the MUSE, OSIRIS, and FLAMES spectra a GC for which spectra from all three instruments are available as illustration of the different datasets. 
In this figure, the different spectral resolutions becomes evident, as well as the imperfect flux calibration of the OSIRIS spectra.

\subsection{Fitting MUSE spectra}
\label{sect:fitting_MUSE_spectra}
\begin{figure}
    \centering
    \includegraphics[width=0.99\linewidth]{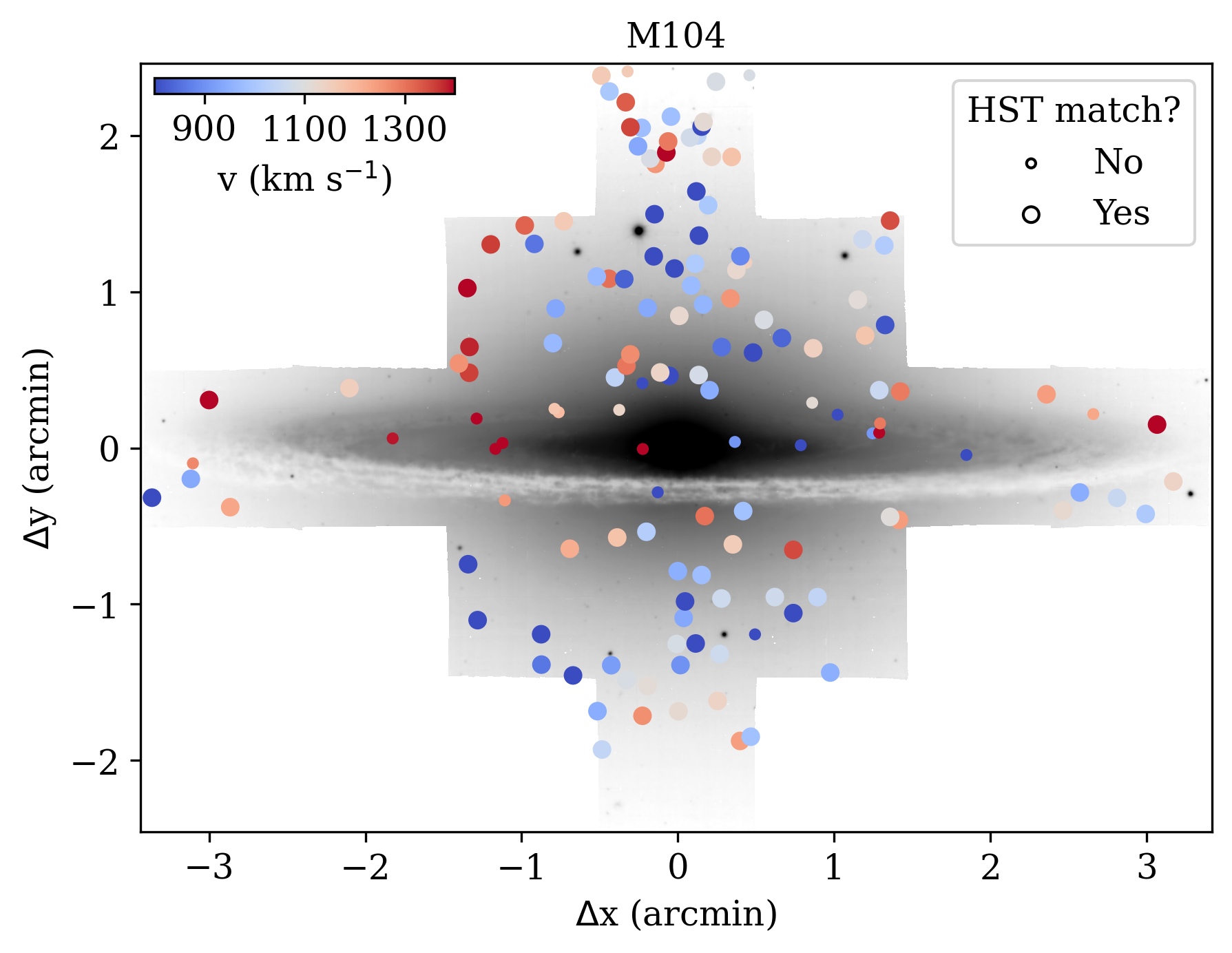}
    \caption{Distribution of GCs extracted from MUSE data. Individual GCs are marked with dots colour-coded according to their LOS velocity. Small dots are GC candidates without match to the HST catalogue from \cite{Spitler2006} and \cite{Harris2010}. The greyscale image shows the white light image from MUSE.}
    \label{fig:MUSE_GCs_vel}
\end{figure}

For fitting the MUSE spectra, we used the E-MILES\footnote{\url{http://research.iac.es/proyecto/miles/pages/spectral-energy-distributions-seds/e-miles.php}} single stellar population (SSP) models \citep{Vazdekis2010, Vazdekis2016}. Those are SSP models based on the empirical MILES stellar spectra, combined with BaSTI isochrones \citep{Pietrinferni2004, Pietrinferni2006} and a double power-law initial mass function with a high mass slope of 1.30 \citep{Vazdekis1996}. The models span an age range between 30 Myr and 14 Gyr and have total metallicities between [M/H] = $-2.27$ dex and $+0.40$ dex. For comparison purposes, we also fitted the spectra with SSPs based on the X-Shooter spectral library (XSL)\footnote{\url{http://xsl.u-strasbg.fr/page_ssp.html}} presented in \cite{Verro2022}. We chose the models based on the PARSEC/COLIBRI isochrones \citep{Bressan2012, Marigo2013, Chen2015} using the initial mass function from \cite{Kroupa2001}. The models have ages above 50 Myr, and iron metallicities between [Fe/H] = $-$2.2 and $+$0.2 dex.

While the E-MILES models reach up to [M/H] = $+0.4$ dex, the XSL SSPs give models up to [Fe/H] = $+0.2$ dex, which artificially decreases the metallicities derived from the XSL models at high metallicities. Additional differences might be caused by light element abundances. The E-MILES models assume [M/H] = [Fe/H]; however, this is only strictly true at high metallicities due to empirical MILES spectra that were used to construct the models that follow the abundance pattern of the solar neighbourhood. The E-MILES models are sampled at 1.25 \AA\,pix$^{-1}$ with a spectral resolution of 2.5 \AA, while the XSL SSPs have a resolution of $R \sim 10,000$ ($\sigma = 16$ km s$^{-1}$). When fitting MUSE spectra, we considered the wavelength-dependent line spread function description from \cite{Guerou2017}.

When fitting the GC candidate spectra, we took a two step approach. Firstly, we fitted for the LOS velocities using additive polynomials of degree 12. Then, all candidates with S/N > 7.5 were fitted with fixed velocities and no additive polynomials. Multiplicative polynomials of degree 8 were used to minimise template mismatch. As \textsc{pPXF} returns the weights of the best-fitting combination of models, we can reconstruct the best-fitting age and metallicity from the fit. Sky residual lines were masked and we only considered the wavelength range between 4700 and 7100 \AA. While we left the velocity dispersion as a free parameter in the fitting, we did not attempt to constrain it as the expected velocity dispersions of GCs ($< 20$ km s$^{-1}$) is significantly lower than the MUSE instrumental resolution ($\sim$ 80 km s$^{-1}$).

After performing initial fits for all 258 GC candidates from the MUSE data, we inspected the fits and removed obvious contaminants such as foreground stars and background galaxies from the sample. The remaining 179 spectra are then fitted again with a Monte Carlo (MC) approach to derive statistical uncertainties on the parameters. Here, the best-fitting spectrum is perturbed using the residual of the initial fit to randomly draw noise in each wavelength. This creates 100 realisations of the spectrum that are then fitted to obtain distributions of the velocity, age, and metallicity.
The reported values refer to the mean and standard deviation of the fits. To speed up the fitting procedure, we restricted the model library to SSPs with ages $> 5$ Gyr, as we do not see any indication of younger ages from the initial fits.

Of the full sample, we kept 152 GC candidates that have velocity uncertainties $<$ 100 km s$^{-1}$, 114 of them are matched to the HST catalogues. Of those, 126 have uncertainties $<$ 50 \mbox{km s$^{-1}$} (98 matched to HST). Most of the GCs without a match to the HST catalogues are in the disc region M\,104 and might therefore be clumps embedded in the dust lanes rather than genuine GCs. To remove those, we inspected all GCs without match to the HST catalogue \citep{Harris2001} on the HST colour image and kept only those with clear cluster-like morphology. This leaves us with a final sample of GCs with MUSE spectra of 139 GCs with velocity uncertainties lower than 100 km s$^{-1}$ (118 with uncertainties smaller than 50 km s$^{-1}$). The MUSE sample is shown in Fig. \ref{fig:MUSE_GCs_vel} on top of a white-light image of the MUSE data. We discuss the completeness of the MUSE sample in Appendix \ref{sect:completeness_MUSE}.

In addition to the GCs, we also used the MUSE data to investigate the stellar light of M\,104. We binned the data cube using a Voronoi binning scheme with the \textsc{vorbin} package \citep{Cappellari2003} to a signal-to-noise of 300, resulting in 626 bins. Those binned spectra were fitted with the same \textsc{pPXF} setup to obtain velocities, velocity dispersions, and stellar population properties. While a full analysis of the stellar component of M\,104 is beyond the scope of this work, we used these properties for comparison against the GC sample.

\subsection{Fitting OSIRIS spectra}
\label{sect:GTC_specs}

To fit the GTC/OSIRIS spectra, we chose the same setup for \textsc{pPXF} as for the MUSE spectra, except for adapting the E-MILES and XSL SSP models to the OSIRIS resolution of FWHM = $\sim 5$ \AA. Due to the decreasing sensitivity in the blue and prominence of sky line residuals in the red, we chose a wavelength range between 4550 and 7100 \AA. 
Before fitting the GTC spectra, we rebinned them to a common wavelength array with a sampling of 2.0 \AA.

As for the MUSE spectra, we first fitted all spectra and then did a visual inspection of the fits to remove five low S/N spectra and one background galaxy. In a second step, we used an MC approach to derive mean velocities, ages, and metallicities for a final sample of 69 GCs.

\subsection{Fitting FLAMES spectra}
\label{sect:FLAMES_spec}
Due to the significantly higher spectral resolution of FLAMES, we used the stellar atmosphere models from \cite{Coelho2014}\footnote{http://specmodels.iag.usp.br/} to obtain radial velocities with \textsc{pPXF}. Those models have a sampling of 0.02 \AA\,pix$^{-1}$ at a spectral resolution of $R = 20,000$, which we match to the FLAMES resolution (FWHM = 0.54 \AA) by convolution with a Gaussian kernel. To fit the velocities of GCs, which are dominated by the light from red giant branch stars in the wavelength range of the FLAMES spectra, we selected stars with $4000< T_{\rm eff}<5000$ and $0.5<\text{log}(g) < 3$. Velocity dispersions obtained from the high resolution FLAMES spectra will be presented in a future work.

Similar to the other datasets, we first fitted all FLAMES spectra to obtain initial estimates of the LOS velocities. Then, we removed those sources with velocities $< 200$ km s$^{-1}$, which are foreground stars rather than GCs belonging to M\,104, and those that did not show absorption lines in the spectrum, leaving 110 GCs.
In the second step, we performed MC fitting with 100 iterations to obtain statistical uncertainties on the velocities. Due to the high resolution of FLAMES, we found typical velocity uncertainties of the order of 0.5 - 2 km s$^{-1}$.

To measure metallicities from the FLAMES spectra, we also fit the spectra with the XSL and E-MILES SSP models. Due to the lower spectral resolution of the models compared with the data, we convolved the FLAMES spectra to match that of the models, and then repeated the fitting to obtain also metallicities in a similar approach as for the MUSE and OSIRIS spectra. We restricted the SSPs to ages $>$ 10 Gyr to limit the age-metallicity degeneracy. As the calcium triplet is rather insensitive to age (e.g. \citealt{SakariWallerstein2016}), we do not expect that this impacts our results, also because there are no indications of young ages from the MUSE or OSIRIS GCs.

\section{Catalogue of GCs around the Sombrero galaxy}
\label{sect:catalogue}
In this section, we describe the final GC samples from the different datasets and compare their velocities and metallicities. The final catalogue is then also described in Appendix \ref{app:catalogue}.

\subsection{Removing contaminants}
\label{sect:contaminants}
\begin{figure}
    \centering
    \includegraphics[width=0.46\textwidth]{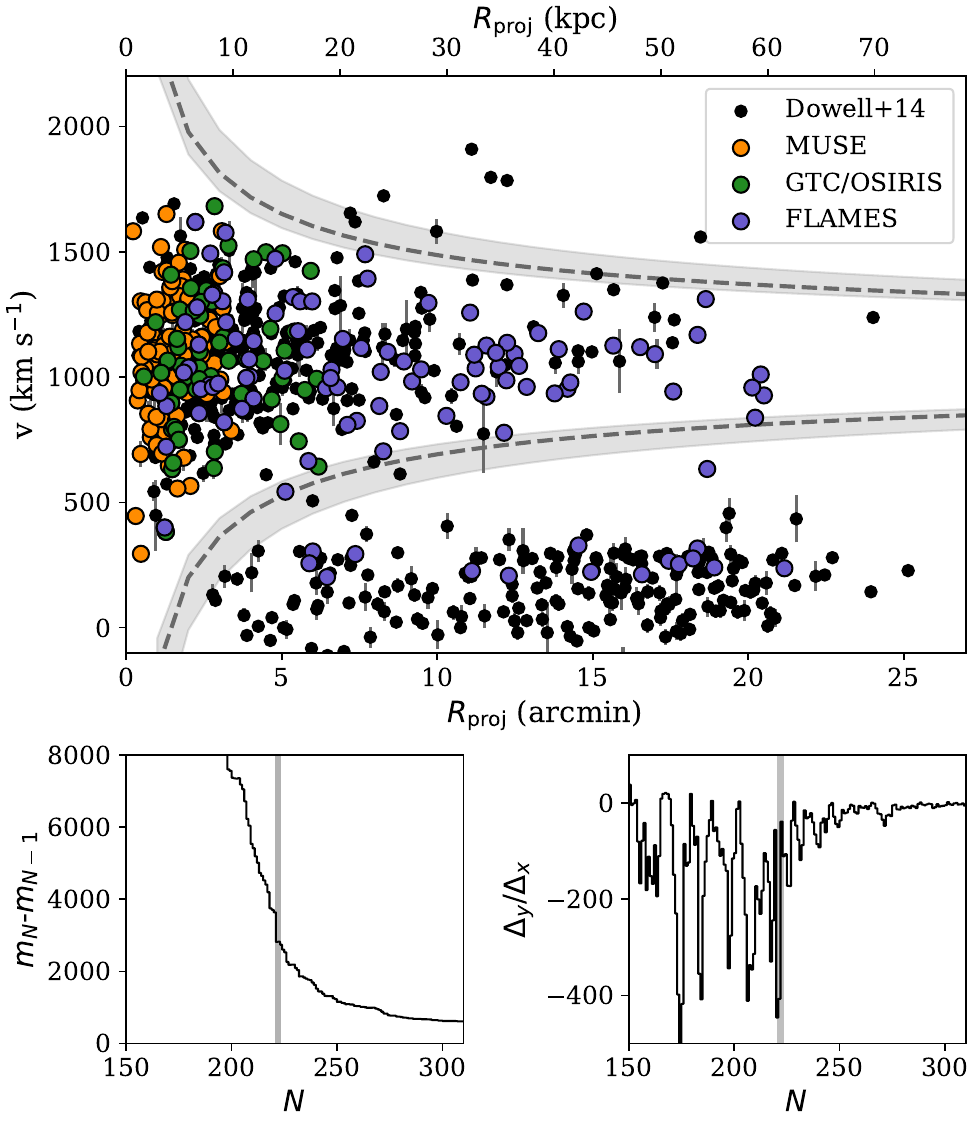}
    \caption{Outlier removal based on GC velocities and projected distances. Top: Line-of-sight velocities as a function of projected galactocentric distance for the full sample of 721 unique sources. Different colours refer to different samples. In case a source is found in multiple samples, only the velocity with the smallest uncertainty is shown. Bottom: Difference between $m_N$ and $m_{N-1}$ as a function of GC number $N$ (left, see Eq. \ref{eq:mn}) and the derivative (right). The vertical grey line indicates our chosen cut-off $N_\text{cut}$ = 222, right after the last big jump as seen in the derivative. From this value, the corresponding $v_\text{max}(R)$ curve was derived, which is shown as the grey dashed line in the top panel. The shaded regions show the curves when assuming $N_\text{cut} \pm 5$.}
    \label{fig:vel_vs_rad}
\end{figure}

\begin{table*}[]
    \centering
    \caption{Overview of GC samples.}
    \begin{tabular}{c c c c c c} \hline\hline
        Sample & N$_{\text{GCs}}$ & <v> & <$\Delta$v> & <$S/N$>  & <R$_\text{gal}$> \\
            & & (km s$^{-1}$) & (km s$^{-1}$) & (pix$^{-1}$) & (arcmin) \\ 
            (1) & (2) & (3) & (4) & (5) & (6) \\ \hline
        MUSE & 139 & 1080.2 & 21.1 & 7.5 &  1.4 \\
        FLAMES & 94 & 1036.4 & 0.9 & 11.4 &  6.9 \\
        OSIRIS & 69 & 1033.7 & 22.6 & 74.4 &  2.6 \\
        \cite{Dowell2014} & 365 & 1101.0 & 15.0 &  ... & 3.8 \\ \hline
    \end{tabular}
    \tablefoot{(1) Name of sample, (2) number of GCs, (3) median velocity, (4) median velocity uncertainty, (5) median S/N per pixel, (6) median galactocentric distance. We note that the catalogue from \cite{Dowell2014} contains also a collection of earlier spectroscopic catalogues from \cite{Bridges1997}, \cite{Larsen2002}, \cite{Bridges2007}, and \cite{AlvesBrito2011}.}
    \label{tab:sample_overview}
\end{table*}

Collecting the objects with successful fits from MUSE, OSIRIS, and FLAMES results in 302 sources. Together with the full catalogue of 579 sources from \cite{Dowell2014}, we have collected 881 radial velocity measurements of objects around M\,104. Before removing contaminants such as foreground stars and unbound GCs, we cross-matched the different catalogues to identify duplicates. Using a matching radius of 1\arcsec, we could identify 721 unique sources. 
Figure \ref{fig:vel_vs_rad} shows the LOS velocities of these sources as a function of the projected distance from the galaxy centre. In case a source is contained in more than one of the samples, we show here the velocity with the smallest uncertainty. As shown in Sect. \ref{sect:comparing_samples}, the velocities between the different samples agree.

As can be seen in Fig. \ref{fig:vel_vs_rad}, the samples span over different regions, with the MUSE sample tracing the innermost GCs and the FLAMES sample reaching out to 20\arcmin. While we already removed obvious contaminants such as background galaxies and foreground stars with $v < 200$ km s$^{-1}$, Fig. \ref{fig:vel_vs_rad} shows that there is a population of foreground stars, as also described in \cite{Dowell2014}, with small LOS velocities, as well as possible intra-group GCs with high velocities. 

To remove contaminants from the sample, we followed the approach described in \cite{Dowell2014}, which is based on the method from \citealt{Schuberth2010, Schuberth2012}. The approach is based on a tracer mass estimator from \cite{Evans2003} that can be applied when the distribution of tracers is different from the overall mass distribution, as for the case of GCs. For this, the quantity $m_N$ of $N$ GCs is calculated as
\begin{equation}
    m_N = \frac{1}{N} \sum_{i = 1}^{N} v_i^2 R_i,
    \label{eq:mn}
\end{equation}
where $v_i$ and $R_i$ are the velocity relative to the systemic velocity and projected radius of the $i$th GC. We assume here a systemic velocity of 1089.1 km s$^{-1}$ as derived by \cite{Sutter2022} from the MUSE data. We verified this velocity with our binned stellar kinematic analysis. $m_N$ is proportional to the tracer mass estimate, but if all GCs including outliers are considered, this estimate is very large. Therefore, to probe where the tracer mass estimates starts to converge, the source with the largest $v_i^2 R_i$ is removed from the sample and $m_N$ is recomputed. This is repeated for all objects, and since the GCs belonging to M\,104 should trace the same underlying mass, $m_N$ should converge when all outliers are removed. Consequently, the cut-off value $N_\text{cut}$ is identified as the point when the curve $m_N$ - $m_{N-1}$ as a function of $N$ begins to flatten (Fig. 
\ref{fig:vel_vs_rad}). As this choice is somewhat subjective, we followed here \cite{Dowell2014} and \cite{Schuberth2010}, and set $N_\text{cut}$ as the point after the last large jump in the $m_N$ - $m_{N-1}$ curve. This jump was identified from the derivative (bottom right panel in Fig. \ref{fig:kin_results_vs_rad}). We used $N_\text{cut} = 222$ to balance between removing clear outliers and still retaining a large GC sample. Other choices that are within $N_\text{cut} = 222 \pm 5$ also seem reasonable, but for lower values, some of the foreground stars would be kept in the sample, and for higher values of $N_\text{cut}$, GCs would be removed.

In the last step of the outlier removal, $C_\text{max}$ = $v_{N_\text{cut} +1}^2 R_{N_\text{cut} +1}$, is derived as the $v^2 R$ value of the first non-removed GC. Based on this, the $v_\text{max}(R) = \sqrt{C_\text{max}/R}$ curve is obtained to select the final GC sample (dashed line in the top panel of Fig. \ref{fig:vel_vs_rad}). Using this selection, we obtained a sample of 499 GCs, which we use in the following. We note that a different choice for $N_\text{cut}$ does not alter the results significantly. Compared to the estimated total number of GCs (1500 - 2000, \citealt{RhodeZepf2004, Kang2022}), this constitutes a significant fraction ($\sim 25 - 30$\%) of all GCs.

\begin{figure*}
    \centering
    \includegraphics[width=0.95\textwidth]{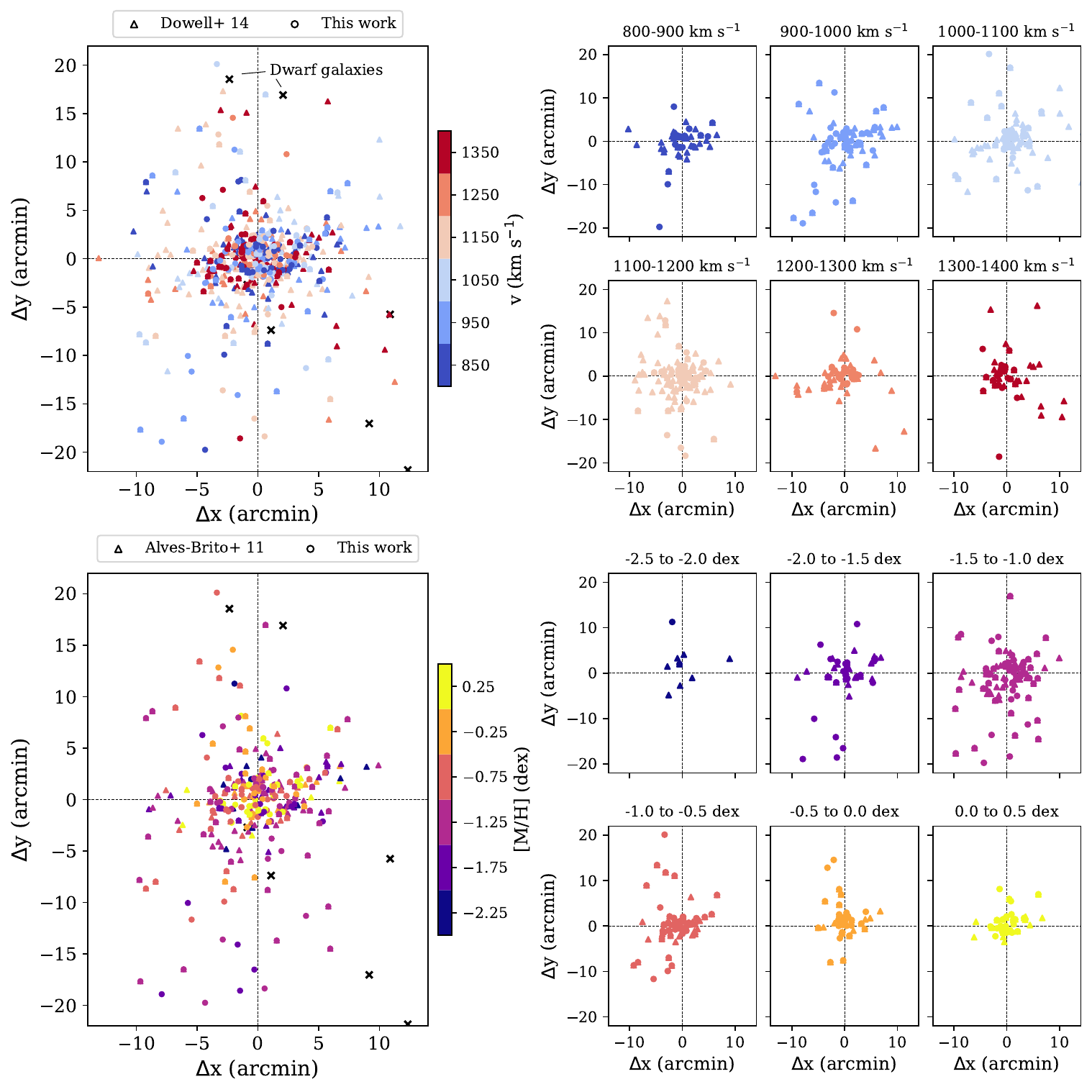}
    \caption{Spatial distribution of GC velocities (top) and metallicities (bottom). 
    The large panels on the left show the full distribution of velocities and metallicities, respectively. Circles refer to this work; triangles refer to GCs listed in the \citealt{Dowell2014} and \citealt{AlvesBrito2011}. Black crosses show dwarf galaxy candidates near M\,104 from \cite{Crosby2024}. The smaller panels on the right side show the velocities and metallicities in different bins, as indicated in the panel titles. Fig. \ref{fig:GC_vel_metal_sky_alt} shows a zoom-in of the central 5\arcmin $\times$ 5\arcmin.}
    \label{fig:GC_vel_metal_sky}
\end{figure*}

To visualise this sample, we show the distribution of 499 GC velocities and 278 metallicities (190 of those were not yet reported in \citealt{AlvesBrito2011}) in Fig. \ref{fig:GC_vel_metal_sky}. In Appendix \ref{app:central_distribution}, we show a zoom into the central 5\arcsec $\times$ 5\arcsec. Due to the two FLAMES pointings north and south of the disc, the distribution is asymmetric, but reaches out to $\sim$ 25\arcmin\,from the centre of M\,104. In addition to GCs, we also added the coordinates of dwarf galaxy candidates near M\,104 marked as crosses \citep{Crosby2024}. Interestingly, one GC from the \cite{Dowell2014} catalogue can be associated with the dwarf galaxy dw1239-1143. Inspecting the imaging presented in \cite{Crosby2024} of this dwarf suggests that the corresponding GC is the nuclear star cluster of this dwarf. Additionally, we note that we covered SUCD1 \citep{Hau2009}, an ultra-compact dwarf (UCD) galaxy in our sample with a FLAMES spectrum. As discussed in \cite{Hau2009}, this object has similar properties as other UCDs or massive GCs. They discussed that its elevated velocity dispersion and brightness might indicate that it is the remnant nucleus of a disrupted dwarf galaxy rather than a dwarf galaxy itself. This interpretation might also be supported by the discovery of a stellar stream around M\,104 that could be connected to this UCD \citep{MartinezDelgado2021}.

Inspecting the spatial distributions of velocities and metallicities in Fig. \ref{fig:GC_vel_metal_sky}, we see that GCs with more extreme relative velocities tend to be centrally concentrated. This is a direct effect of the gravitational potential of M\,104, which dictates the escape velocity, as also visible in the declining $v_\text{max}(R)$ curve in Fig. \ref{fig:vel_vs_rad}. While overall, no clear rotation signature is visible, there might be a trend of higher relative velocities to the east side of the centre than to the west within the inner $\sim$ 3\arcmin, which would follow the rotation of the disc. The most metal-rich GCs are found close to the centre, while lower metallicities dominate at larger separations, as is further discussed in Sect. \ref{sect:results}. A small population of extremely metal-poor GCs with [M/H] $< -2$ dex is found to be also centrally concentrated, but the low numbers hinder any conclusion about their spatial distribution.

\begin{figure*}
    \centering
    \includegraphics[width=0.85\textwidth]{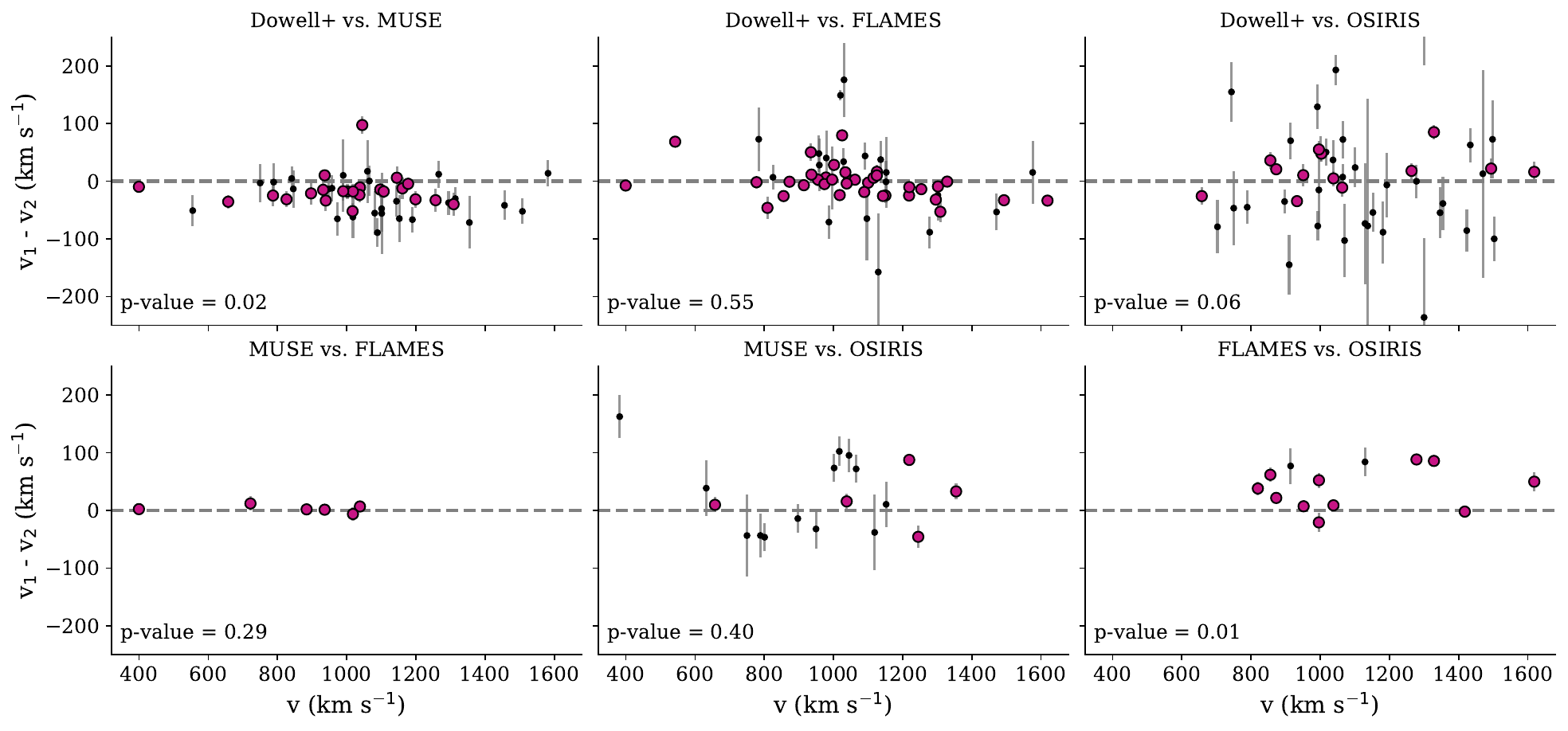}
    \caption{Comparison of LOS velocities. Black dots are all the GCs found in both samples as indicated by the titles of the panels. Large pink circles are GCs where the velocity uncertainty in both samples is less than 20 km s$^{-1}$. In the bottom-left corner, we report the p-value of a paired t-test.}
    \label{fig:GC_velocity_comparison}
\end{figure*}

\subsection{Comparing GC samples}
\label{sect:comparing_samples}

In the final sample of 499 GCs, 365 were listed by \cite{Dowell2014}, 139 have velocities from MUSE, 94 from FLAMES, and 69 from OSIRIS. Table \ref{tab:sample_overview} gives an overview of the different samples, including the numbers, median velocity uncertainties, and $S/N$ values. As can be seen, the high resolution FLAMES spectra give by far the most precise velocity measurements. Due to the large field of view of FLAMES, this sample also probes the largest galactocentric distances, whereas MUSE allowed us to probe the central regions.

In Fig. \ref{fig:GC_velocity_comparison}, we compare the velocities of the different samples using the GCs that are listed in more than one catalogue. The large dots in the figure refer to GCs with velocity uncertainties $< 20$ km s$^{-1}$ in the corresponding catalogues and we use those to test whether there are significant offsets between the samples using paired t-tests. The p-values are reported in the bottom left corners of the panels in Fig. \ref{fig:GC_velocity_comparison}. 
We find no significant offsets with p-values $>$ 0.05 in most cases, except when comparing the catalogue from \cite{Dowell2014} with MUSE and when comparing FLAMES and OSIRIS measurements.
In the latter case, the comparison is only based on a few GCs with a large scatter in the velocity differences, which makes it difficult to assess if there is a systematic offset, especially since the two instruments have very different velocity resolutions. The mean difference between the MUSE and the matched velocities from \cite{Dowell2014} is $<v_\text{Dowell+14} - v_\text{MUSE}> = 15.2$ km s$^{-1}$ with a standard error of the mean of 5.9 km s$^{-1}$. We chose not to apply this offset because the comparisons between the heterogenous \cite{Dowell2014} sample and velocities from FLAMES and OSIRIS do not show a similar consistent offset. Additionally, the few matches between MUSE and FLAMES show a very good agreement ($<v_\text{MUSE} - v_\text{FLAMES}> = 3.0$ km s$^{-1}$ with a standard error of 2.5 km s$^{-1}$). However, we note that applying such an offset to the MUSE velocities (or the velocities from \citealt{Dowell2014}) does not significantly affect our results.

\subsection{Comparing metallicities}
\label{sect:comparing_metallicities}
\begin{figure*}
    \centering
    \includegraphics[width=0.85\textwidth]{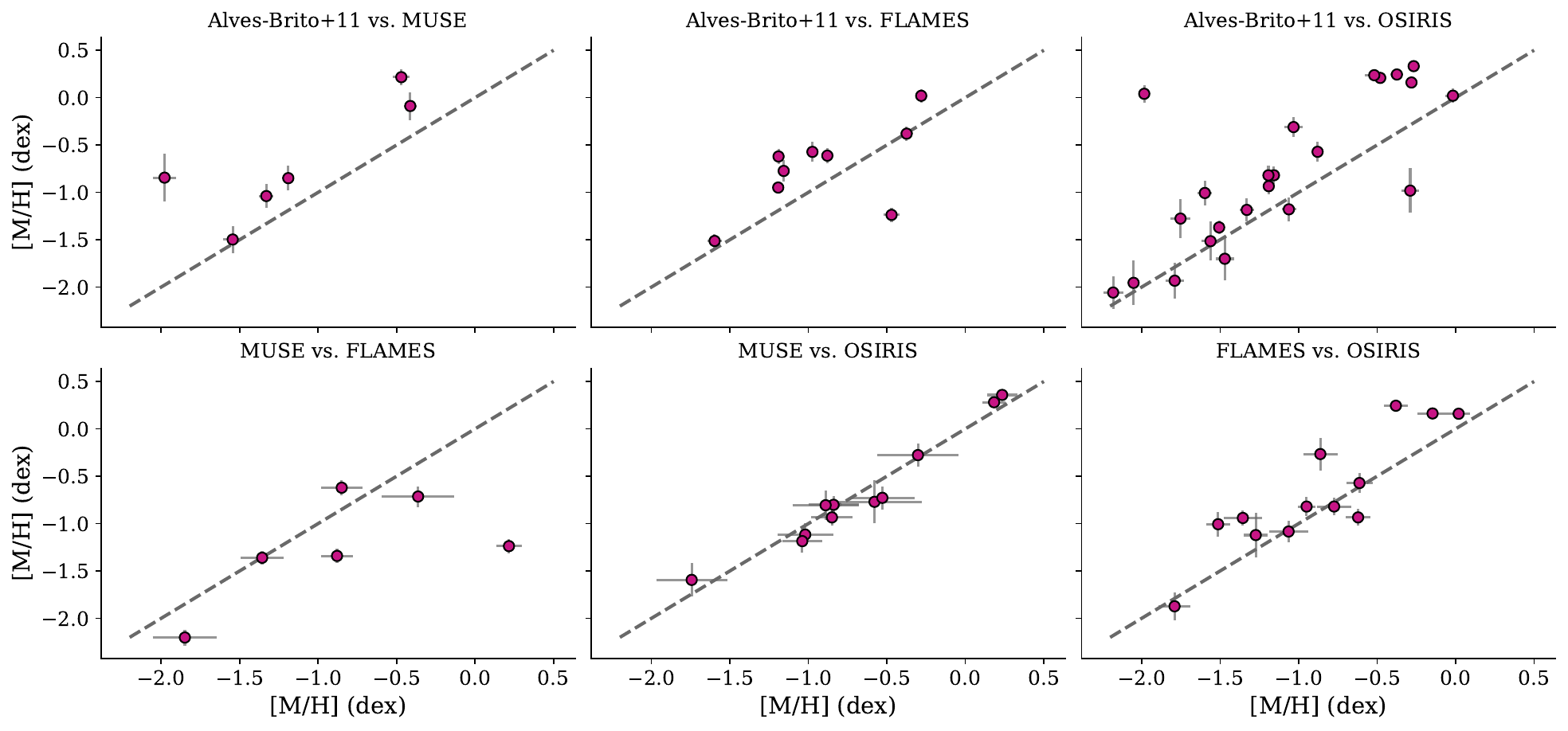}
    \caption{Comparison of GC metallicities. The first sample reference in the panel title refers to the x-axis and the second to the y-axis. The grey dashed line is the one-to-one relation. We note that the literature metallicities from \cite{AlvesBrito2011} refer to iron metallicities, while we show here the total metallicities inferred from the E-MILES SSP models.}
    \label{fig:GC_metallicity_comparison}
\end{figure*}

In Fig. \ref{fig:GC_metallicity_comparison}, we compare the metallicities of GCs listed in at least two of the samples. For the MUSE and OSIRIS samples, we used here the total metallicities as obtained with the E-MILES SSP models. As a literature comparison, we used the mean metallicities from \cite{AlvesBrito2011}, who used line index measurements to estimate iron metallicities. While there can be differences between iron and total metallicities, for example depending on the light element abundance ratio, we show in Appendix \ref{app:EMILES_vs_XSL} that the iron metallicities obtained when fitting with the XSL SSPs agree very well with the total metallicities from the E-MILES models. However, as the XSL SSPs only reach up to $+0.2$ dex, we chose to use the total metallicities from the E-MILES SSPs, as many of the GCs reach high metallicities. 

Overall, the metallicities obtained here agree reasonably well with the results from \cite{AlvesBrito2011}, even though we used full spectrum fitting. In addition, the metallicities from the different instruments (MUSE, FLAMES, OSIRIS), also agree with each other. The best match is found between the MUSE and OSIRIS metallicities, which is expected because both instruments cover the same optical wavelength range and the same fitting approach was taken. Interestingly, also the metallicities obtained from fitting the narrow calcium triplet region of the FLAMES spectra agree with the optical measurements, as especially the comparison between FLAMES and OSIRIS metallicities shows. While there seems to be indications that the metallicities from FLAMES might saturate at the highest metallicities ([M/H] > 0 dex, see also \citealt{Usher2015, Sakari2016, Chung2016}), this is encouraging because it allowed us to trust the FLAMES metallicities at larger galactocentric distances, where no optical spectra are available and GC metallicities are expected to be lower than in the central region (see Sect. \ref{sect:radial_metallicity}).

\section{Kinematics and metallicities of the GC system of M\,104}
\label{sect:results}

In the following section, we use our collected catalogue of GC velocities and metallicities for an initial exploration of M\,104's GC system. We focus on a simple kinematic model describing the rotation and velocity dispersion of the GC system and investigate the GC metallicity distribution of M\,104.

\subsection{Simple kinematic model}
\label{sect:kin_model}
We modelled the kinematics of the GC system with a simple sinusoidal model to constrain the rotation and velocity dispersion of the GC system as a function of radius from the galaxy centre. We followed the approach described in \cite{Fahrion2020b} and \cite{Fahrion2025}, which is similar to the method described in \cite{Dowell2014} for M\,104. In this sinusoidal model, the rotation of the GC system is described as
\begin{equation}
    v_\text{GC,i} (\theta) = v_\text{sys} + v_\text{rot}\,\text{sin}(\theta_i - \theta_0),
\end{equation}
where $v_\text{rot}$ describes the rotation amplitude around the systemic velocity $v_\text{sys}$ with $\theta_0$ denoting the angle of the rotation axis (maximum rotation is found at $\theta_0 + 90^{\circ}$). $v_\text{GC,i}$ is the velocity of the $i$th GC at position angle $\theta_i$, which is used as input to the model. In this model, only the global rotation and dispersion is considered, and any change with radius has to be implemented by choosing GCs at different radii.

The likelihood function is then given as
\begin{equation}
\mathcal{L} = \prod \limits_{i} \frac{1}{\sqrt{2 \pi \sigma_\text{obs}^2}} \text{exp} \left(- \frac{(v_\text{GC, i} - (v_{\text{sys}} + v_\text{rot}\,\text{sin}(\theta_i - \theta_\text{0})))^2}{2 \sigma_\text{obs}^2} \right),
\label{eq:kin_model}
\end{equation}
where $\sigma^2_{\text{obs}} = (\delta v_\text{GC,i})^2 + (\sigma_\text{GCS})^2$
describes the velocity dispersion $\sigma_\text{GCS}$ under the assumption that the observed dispersion has contributions from the intrinsic GC system dispersion and the velocity uncertainty $\delta v_\text{GC,i}$ of the $i$th GC. 

We used the python module \textsc{emcee}\footnote{\url{https://emcee.readthedocs.io/en/stable/}} \citep{emcee} to implement the model using a Markov chain Monte Carlo sampler in a Bayesian framework. As described in Sect. \ref{sect:contaminants}, we used the value with the lowest velocity uncertainty for GCs listed in two or more of the samples. When modelling the full GC system using all 499 GC velocities, we used flat priors on all parameters. For the rotation amplitude, we allowed positive values up to 400 km s$^{-1}$ and constrained the position angle between $-180^\circ$ and $180^\circ$. The velocity dispersion prior ranges between 0 and 500 km s$^{-1}$ and for the systemic velocity, we used a range between 800 and 1500 km s$^{-1}$. For the full GC system, we find a dispersion of $\sigma_\text{GCS} = 229.0^{+7.5}_{-7.5}$  km s$^{-1}$, a low rotation amplitude of $v_\text{rot} = 35.8^{+15.4}_{-15.9}$ km s$^{-1}$, a position angle of $\theta = 21.2^{+25.1}_{-23.9}$ degrees, and a systemic velocity of $v_\text{sys} = 1087.9^{+10.3}_{-10.2}$ km s$^{-1}$, very similar to the systemic velocity of the galaxy as reported in \citealt{Sutter2022} based optical emission lines in the MUSE data.

\begin{figure}
    \centering
    \includegraphics[width=0.49\textwidth]{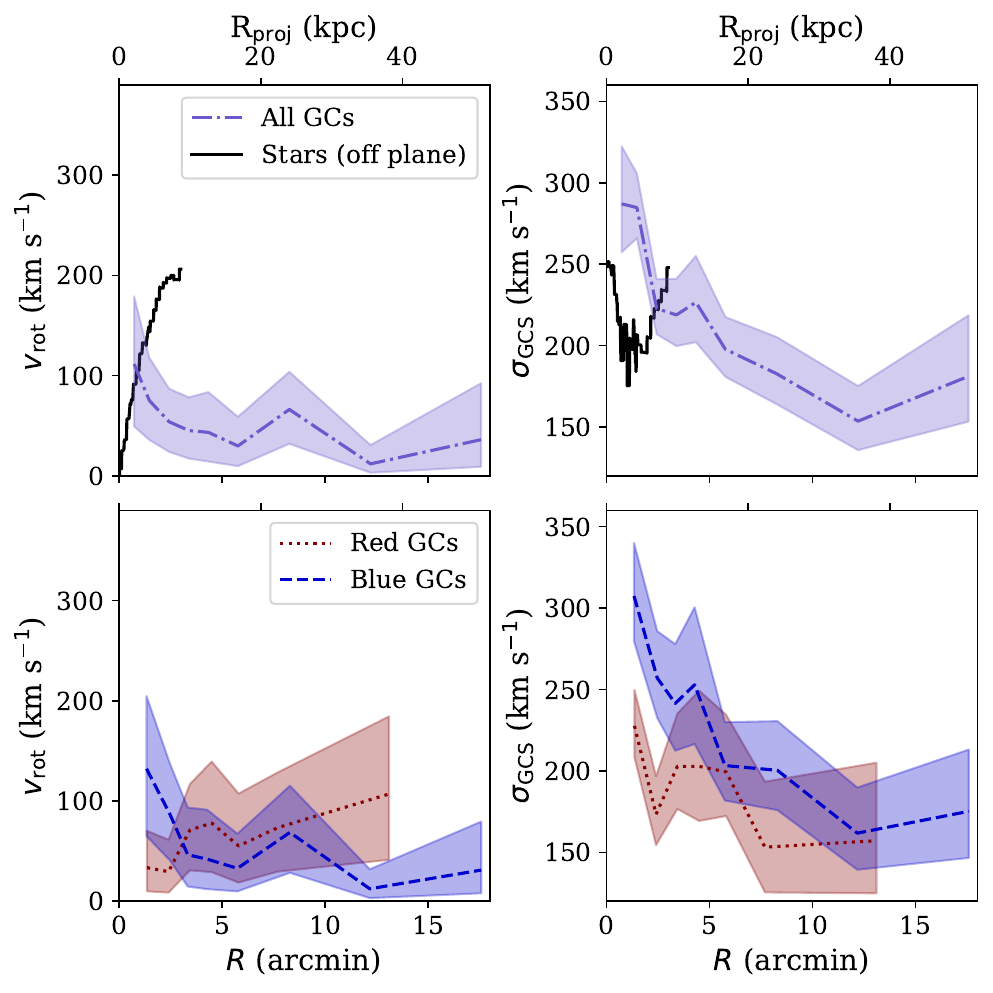}
    \caption{Results of the simple kinematic model as a function of radius. Top left: GC system rotation amplitude $v_\text{rot}$ for all GCs in purple. The black line shows the LOS velocities of the stellar component, taken along a slit 0.4\arcmin\,above the plane of the disc. Top right: Velocity dispersion of the full GC sample $\sigma_\text{GCS}$ in purple. Within the rotation-dominated disc, the dispersion is significantly lower than off-plane. Bottom panels: GC rotation amplitude and dispersion for the red and blue GC subpopulations, separately.}
    \label{fig:kin_results_vs_rad}
\end{figure}

Due to the large number of GCs available, we were able to constrain the parameters at different radii, choosing radial bins out to 18\arcmin. The resulting profiles of rotation amplitude and dispersion are shown in the top panel of Fig. \ref{fig:kin_results_vs_rad}. We fixed the systemic velocity to the systemic velocity of the galaxy ($v_\text{sys} = 1089.1$ km s$^{-1}$, \citealt{Sutter2022}) to avoid effects from the inhomogeneous spatial coverage of the velocity measurements at larger radii. For the same reason, we also constrained the position angle with a Gaussian prior using a mean of $0^{\circ}$ and a standard deviation of $10^{\circ}$ to fix the angle of maximum rotation to a similar angle as the disc ($\sim 90^{\circ}$). 

In addition to fitting the full sample of GCs, we separated them into blue and red GCs. For GCs with $B, V$, and $R$ magnitudes available (from \citealt{RhodeZepf2004} and \citealt{Harris2010}), we used a separation in colour of $B-V = 0.8$ mag. For GCs with $g - i$ colours from \cite{Kang2022}, we used a separation of $g - i = 1.0$ mag. We adapted the radial bins to ensure that at least ten GCs are in each radial bin. The rotation velocity and dispersion of the red and blue GC subpopulations are shown in the bottom panels of Fig. \ref{fig:kin_results_vs_rad}.

Similar to \cite{Dowell2014}, we find low rotation velocities except for the innermost bin (< 1\arcmin) where the rotation velocity reaches above 100 km s$^{-1}$. To compare this with the stellar component, we took an average LOS velocity profile along a slit 0.4\arcmin\,above the disc shown as the black line in the top left panel of Fig. \ref{fig:kin_results_vs_rad}. We find that the disc itself reaches velocities $\pm 350$ km s$^{-1}$ at low velocity dispersions of $\sigma_\ast \sim 60$ km s$^{-1}$ (see also \citealt{Sutter2022} for a description of the gas kinematics within the disc), so this off-plane slit was chosen to trace the spheroid component of M\,104. In this central, off-plane region covered by the MUSE data, the stars show also high dispersions between 200 -- 250 km s$^{-1}$, while the GCs show even higher dispersions of $\sigma_\text{GCS} = 280.3^{+36.0}_{-29.0}$ km s$^{-1}$. At larger radii, the GC system velocity dispersion drops to $\sigma_\text{GCS} = 152.9^{+34.5}_{-25.3}$ km s$^{-1}$ in the outermost bin ($10 < R < 30\arcmin$, $\sim$ 42 kpc). 

Using the LOS velocities of nine dwarf galaxies within 420 kpc of M\,104, \cite{Crosby2025} find a comparable velocity dispersion of $\sigma$ = 138 $\pm$ 36 km s$^{-1}$. Using 15 dwarfs, some of which reach out to $\sim$ 1 Mpc, \cite{Karachentsev2020} find a dispersion of $\sigma = 204$ km s$^{-1}$, possibly indicating that the velocity dispersion profile rises again at larger radii when the potential of the group becomes dominant. Alternatively, dwarfs at very large separations might no longer be bound to the system. 

Splitting the GC sample into red and blue subpopulations shows that the red GCs show very little rotation within the central bins, but the rotation might pick up at larger separations. In contrast, the blue GCs show larger rotation velocities and dispersions in the central region. Over all radii, the red GCs have lower velocity dispersions then the blue subpopulation, and within the region covered by MUSE, the dispersion profile of the red GCs closely follows that of the stars. Also in other galaxies, it is often found that the red GCs closely follow the stellar kinematics, whereas the blue GCs are more dominated by random motions (e.g. \citealt{Schuberth2010, Pota2013, Fahrion2019}). With the higher dispersion for the blue GCs in M\,104, we see here the same trend, except for the very inner regions where the blue GCs seem to be rotating while the red ones do not. We note that also the stellar rotation in the innermost region does not yet reach large amplitudes, so this low rotation in the red population is perhaps not too surprising, especially since the rotation amplitude of the red GCs then rises with radius. 

In future work, we plan to explore the dynamics of M\,104 and its GC system with a more rigorous dynamical model to provide also an updated measurement of the enclosed mass. In this paper, we only employ the simple mass estimator from \cite{Errani2018} to provide an estimate of the dynamical mass of M\,104 within 1.8 effective radii based on the GCs:
\begin{equation}
    M_\text{dyn} (< 1.8 R_\text{eff, GCS}) = 6.3\, R_\text{eff} \,\sigma_{GCS}^{2} \,G^{-1}.
\end{equation}
Using the effective radius of the GC system of $R_\text{eff, GCS}$ = 7.59 $\pm$ 0.72 \arcmin (21 $\pm$ 2 kpc; \citealt{Kang2022}) and the velocity dispersion when modelling the full system, we derived a mass estimate of $M_\text{dyn} (< 1.8 R_\text{eff, GCS}$) = (1.6 $\pm$ 0.2)$\times$ $10^{12} M_\sun$. Despite being only a simple estimate, this value is in good agreement with the Jeans model presented in \cite{Dowell2014}. At an enclosed radius of 41 kpc (1.95 effective radii of the GC system), they found a total mass of $M_\text{dyn} (< 41\,\text{kpc}) = 1.3 \times 10^{12} M_\sun$.

\subsection{Radial metallicity profile}
\label{sect:radial_metallicity}
\begin{figure}
    \centering
    \includegraphics[width=0.48\textwidth]{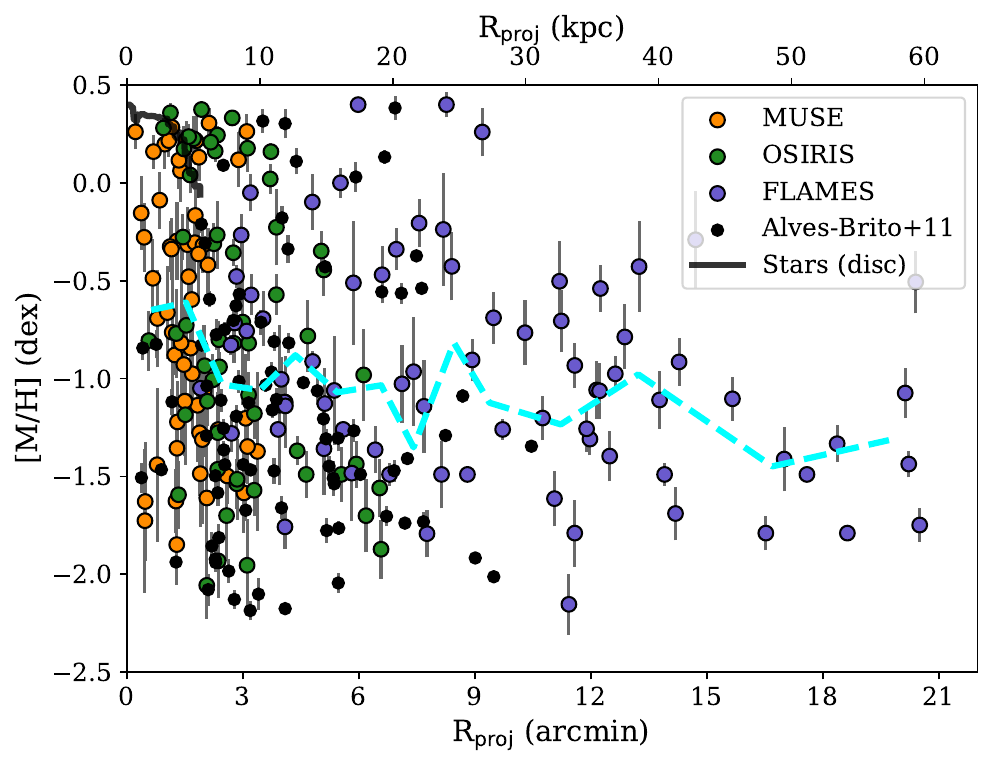}
    \caption{Globular cluster metallicities as a function of galactocentric distance. Different coloured dots indicate metallicities from different instruments. Smaller back dots refer to metallicities from \cite{AlvesBrito2011}. In case multiple metallicities are available, preference is given to OSIRIS over MUSE over FLAMES. The cyan dashed line shows the mean metallicity profile. The black line shows the metallicity of the stellar disc.}
    \label{fig:radial_metallicity}
\end{figure}

We show the GC metallicities as a function of galactocentric distance in Fig. \ref{fig:radial_metallicity}. For GCs that are in multiple samples we prefered OSIRIS metallicities over MUSE over FLAMES over the line-index based metallicities from \cite{AlvesBrito2011} due to the high $S/N$ in the OSIRIS spectra.
As this figure shows, the GCs of M\,104 cover a large range of metallicities at all radii. In the central region covered by MUSE, values from [M/H] = $-2$ dex to supersolar metallicities are found and this continues for larger radii. We note here that there are GCs that hit the edges of the metallicity grids of the E-MILES SSPs, so their true metallicities likely are even higher. The cyan dashed line in Fig. \ref{fig:radial_metallicity} shows the mean metallicity profile, indicating that the mean metallicity shows a mild negative trend. In addition, we show with a black line the metallicity of the galaxy in a slit along the major axis of the disc, where the highest metallicities are reached.

Within one effective radius ($R_\text{eff}$ = 1.7\arcmin = 4.9 kpc, \citealt{KormendyWestpfahl1989})\footnote{We note that the effective radius of M\,104 has been measured several times using different approaches (e.g. \citealt{GadottiSanchezJanssen2012}) with varying results depending on whether the disc and bulge are considered separately. We used the same value as \cite{Dowell2014}.}, we find a mean GC metallicity of [M/H] $\sim -0.7$ dex, which decreases to $\sim -1$ dex above 3\arcmin ($\sim 2\,R_\text{eff}$) and stays relatively flat out to $\sim$ 12\arcmin ($\sim 7\,R_\text{eff}$).
At larger radii ($> 12$\arcmin), the mean metallicity then drops to [M/H] $\sim -1.3$ dex. 
The stellar metallicity within the disc also shows a gradient towards lower metallicities at larger separations, but overall the stellar metallicity is higher than that of most GCs (e.g. see also \citealt{Lamers2017}). Nonetheless, there are a few GCs as metal-rich as the galaxy disc.

While we only have a few metallicities from FLAMES available at large separations, the decrease in the mean GC metallicity with radius is expected as also the mean colour of GCs decreases with radius \citep{Kang2022}. On the other hand, the rather flat profile and the presence of metal-rich GCs out to large separations is observed around other massive galaxies \citep{Usher2012, Pastorello2014}. For example, \citealt{Villaume2020} reported a flat mean metallicity profile for GCs around M\,87 out to 8 effective radii with a large scatter of metallicity at all radii, similar to our findings.

\subsection{GC metallicity distribution}
\label{sect:metallicity_distribution}

\begin{figure*}
    \centering
    \includegraphics[width=0.95\textwidth]{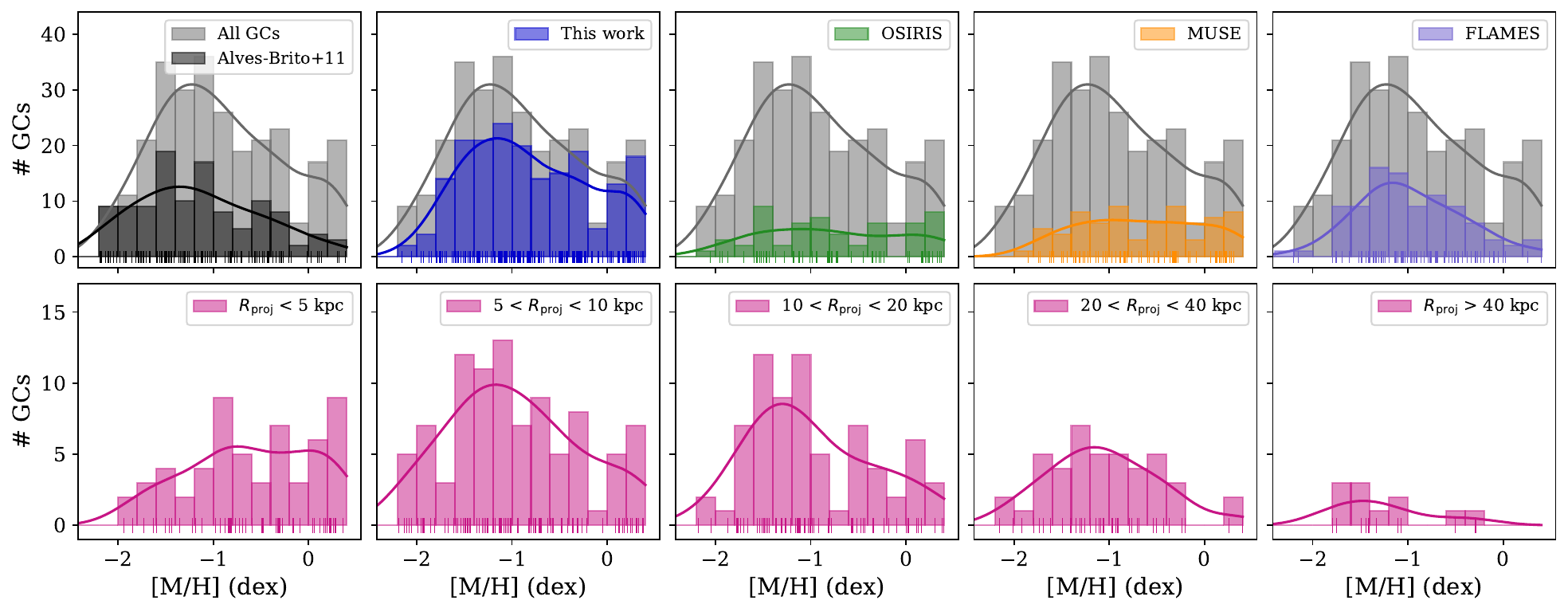}
    \caption{Distribution of GC metallicities. Top row: Full sample of GC metallicities (shown in grey) combining the measurements from \cite{AlvesBrito2011} (black, first panel) and the ones presented here (blue, second panel). Third to fifth panel: GC metallicities from OSIRIS (green), MUSE (orange), and FLAMES (purple). Bottom: GC metallicities in different radial bins. The lines show kernel density estimations scaled to match the counts. The small vertical lines near zero indicate the individual metallicity measurements.}
    \label{fig:metallicity_histograms}
\end{figure*}

We show the GC metallicities as binned histograms in Fig. \ref{fig:metallicity_histograms}. In top panels, we show the full sample of GC metallicities in the grey histograms. As described before, we chose OSIRIS metallicities where available over MUSE over FLAMES over the literature sample of \cite{AlvesBrito2011}. 
Based on index-based measurements, \cite{AlvesBrito2011} suggested that the GC metallicity distribution of M\,104 is bimodal with peaks at [Fe/H] = $-1.44$ and $-0.66$ dex (black histogram in the top left panel). However, we cannot confirm this from our sample as shown in Fig. \ref{fig:metallicity_histograms} in the blue histogram in the second panel. Instead, we find a metallicity distribution that is dominated by a broad peak at \mbox{[M/H] = $\sim$ $-$1.1 dex} and a smaller population of metal-rich GCs. A two-sample Kolmogorov-Smirnov test comparing our sample and that of \cite{AlvesBrito2011} finds a p-value < 0.05, suggesting different distributions, likely because of the different spatial distributions between the samples. Our sample contains many central, metal-rich GCs in the MUSE and OSIRIS samples, which are not as prominent in the sample from \cite{AlvesBrito2011}. On the other hand, the FLAMES sample contains many more metal-poor GCs, likely because the FLAMES footprint covers more of the outer GC system. 

To explore this in more detail, the bottom panels of Fig. \ref{fig:metallicity_histograms} show the GC metallicities in different radial bins, from the centre out to $\sim 70$ kpc (see also Fig. \ref{fig:radial_metallicity}). 
In these histograms, the decreasing mean metallicity becomes evident, as fraction of metal-poor GCs increases with increasing radius. We caution here again that this decrease might also be influenced by a possible saturation of the FLAMES metallicity measurement from the calcium triplet region \citep{Usher2012}. 
Nevertheless, a decrease in mean metallicity is expected at larger radii due to the increasing fraction of accreted material stemming from metal-poor dwarf galaxies (e.g. \citealt{ForbesRemus2018}).

\subsection{The formation history of M\,104}
With their old ages, GCs are often regarded as powerful probes of galaxy evolution. In the traditional picture, metal-poor GCs are thought to originate from accreted low-mass, metal-poor systems. However, since metal-poor GCs can also form in-situ in the host galaxy, either at earlier times or in less enriched regions, assigning individual GCs to either population is a complex problem as simulations suggest \citep{ForbesRemus2018, Pfeffer2023}. Likely only GCs with extreme metallicities -- either very metal-rich or very metal-poor -- can be assigned to either category with confidence.

Comparing the stellar metallicity from the MUSE data (see Fig. \ref{fig:radial_metallicity}) shows that M\,104 is a metal-rich galaxy, as also found in previous studies \citep{MouldSpitler2010}. 
While the MUSE data only probes the inner regions, \cite{Cohen2020} used resolved stellar populations from HST imaging out to almost 40 kpc to study the stellar metallicity. They reported only a negligible fraction of stars below [Z/H] $< -1$ dex. Consequently, metal-rich GCs appear to be good tracers of this metal-rich component out to several effective radii, as was also found for other massive galaxies (e.g. \citealt{Beasley2008, Peacock2015, Harris2016, Fahrion2020b, Villaume2020}). Another indication of this is seen in the different kinematics between the red and blue GCs (Fig. \ref{fig:kin_results_vs_rad}), where the red GCs more closely follow the kinematics of the galaxy. Consequently, there are indications that the red GCs are good tracers of the metal-rich spheroid of M\,104, which might have been built in-situ of from the accretion of massive, metal-rich satellites. For example, \cite{Cohen2020} proposed that the high metallicity in the far halo of M\,104 could be explained by a recent major merger with another massive galaxy ($M_\ast \sim 10^{11}$ $M_\sun$). Such a massive galaxy would then also bring in GCs with a broad range of metallicities.

On the other hand, the high velocity dispersion and low rotation at large radii of the blue GCs suggest that at least some of them they were accreted from lower mass systems, building a dynamically hot halo component. The presence of a large range of GC metallicities, including very metal-poor ones, together with an extended number density profile of blue GCs \citep{Kang2022} implies a rich formation history of M\,104 including also minor mergers. 

\section{Conclusions}
\label{sect:conclusions}

In this work, we have presented a new spectroscopic GC catalogue of 499 GCs based on previous works \citep{Dowell2014, AlvesBrito2011} and new spectra from OSIRIS, MUSE, and FLAMES.
We summarise our main findings as follows:
\begin{itemize}
    \item {
Extracting GC spectra from MUSE integral-field spectroscopy allowed us to probe the inner GC system of M\,104. From the MUSE data, we extracted the spectra of 139 GCs, including a few that are located within the disc. In addition, 69 GCs were observed with OSIRIS, yielding high S/N spectra, and 94 GCs have high spectral resolution spectra from FLAMES.}
\item {Combined with the GC catalogue presented by \cite{Dowell2014}, we obtained a GC sample of 499 GCs around M\,104. Given that the total number of GCs is estimated at 1600 - 2000, this spectroscopic GC sample constitutes a significant fraction ($\sim$ 25 - 30\%) of all GCs of M\,104.}
\item {We find good agreement among the velocity measurements of GCs present in multiple samples. Similarly, when comparing the metallicity measurements, we found excellent agreement between metallicities measured from MUSE and OSIRIS. The metallicities obtained by using full spectrum fitting to the FLAMES calcium triplet region spectra show good agreement with other estimates, highlighting the potential of using those spectra for stellar population studies.}
\item{We modelled the GC velocities with a simple sinusoidal model, and the model suggests that the velocity dispersion decreases with radius. The rotation amplitude is low, illustrating that the GC system is dispersion-dominated, and it traces the enclosed mass, which is also consistent with previous studies. When separating red and blue GCs, we found that the red GCs have lower dispersions and more closely follow the stellar light, while the blue GCs show rotation only in the very inner region and are more dispersion-dominated. A more detailed discrete dynamical model will be subject of a future project.}
\item {Based on metallicity measurements of 278 GCs (190 from MUSE, OSIRIS, and FLAMES), we find that the GC system of M\,104 shows a large scatter in metallicity, and this scatter persists out to the halo region. Within one effective radius ($R_\text{eff} = 1.7$ \arcmin = 4.9 kpc), the mean GC metallicity is [M/H] = $-$0.7 dex, while it decreases to $\sim -1$ dex beyond 3 $R_\text{eff}$ (15 kpc) and remains at that level out to $\sim 50$ kpc. The metal-rich GCs are as metal-rich as the disc of M\,104.}
\item {We find a complex GC metallicity distribution from the spectroscopic measurements. The total distribution has a broad metal-poor peak at [M/H] = $-1$ dex and a smaller population of more metal-rich GCs. Exploring the distribution as a function of projected distance shows broad GC metallicity distributions at all radii with a mild trend towards lower metallicities at larger radii, as also seen in the radial metallicity profile. The presence of metal-rich GCs are found at all radii indicates a metal-rich halo, while the broad range of GC metallicities is likely a consequence of the rich formation history of M\,104.}
\end{itemize}

\section{Data availability}
Table \ref{app:catalogue} is only available in electronic form at the CDS via anonymous ftp to \url{cdsarc.u-strasbg.fr} (130.79.128.5) or via \url{http://cdsweb.u-strasbg.fr/cgi-bin/qcat?J/A+A/}.

\begin{acknowledgements}
We thank the referee for a constructive report that has helped to improve this manuscript.
KF acknowledges funding from the European Union’s Horizon 2020 research and innovation programme under the Marie Sk\l{}odowska-Curie grant agreement No 101103830. O.M. is grateful to the Swiss National Science Foundation for financial support under the grant number PZ00P2\_202104.
This work made use of Astropy:\footnote{\url{http://www.astropy.org}} a community-developed core Python package and an ecosystem of tools and resources for astronomy \citep{astropy2013, astropy2018, astropy2022}. Based on observations collected at the European Southern Observatory under ESO programmes 60.A-9303 and 114.274W, and at the Gran Telescopio Canarias under programmes GTC24-19A and GTC89-20A. ACS acknowledges support from FAPERGS (grants 23/2551-0001832-2 and 24/2551-0001548-5), CNPq (grants 314301/2021-6, 312940/2025-4, 445231/2024-6, and 404233/2024-4), and CAPES (grant 88887.004427/2024-00).
\end{acknowledgements}

\bibliographystyle{aa} 
\bibliography{references}

\appendix

\section{Completeness of MUSE data}
\label{sect:completeness_MUSE}

While multi-object spectroscopy has long been used to derive spectroscopic GC catalogues around galaxies, using MUSE integral field spectroscopy to probe the more central regions is a more recent approach (e.g. \citealt{Fahrion2019, Fahrion2020b, Grasser2024}). To illustrate the capabilities this approach offers in obtaining high quality spectra even in the crowded and bright inner regions of massive galaxies, we briefly explore here the completeness of the MUSE data.

\begin{figure}
    \centering
    \includegraphics[width=0.46\textwidth]{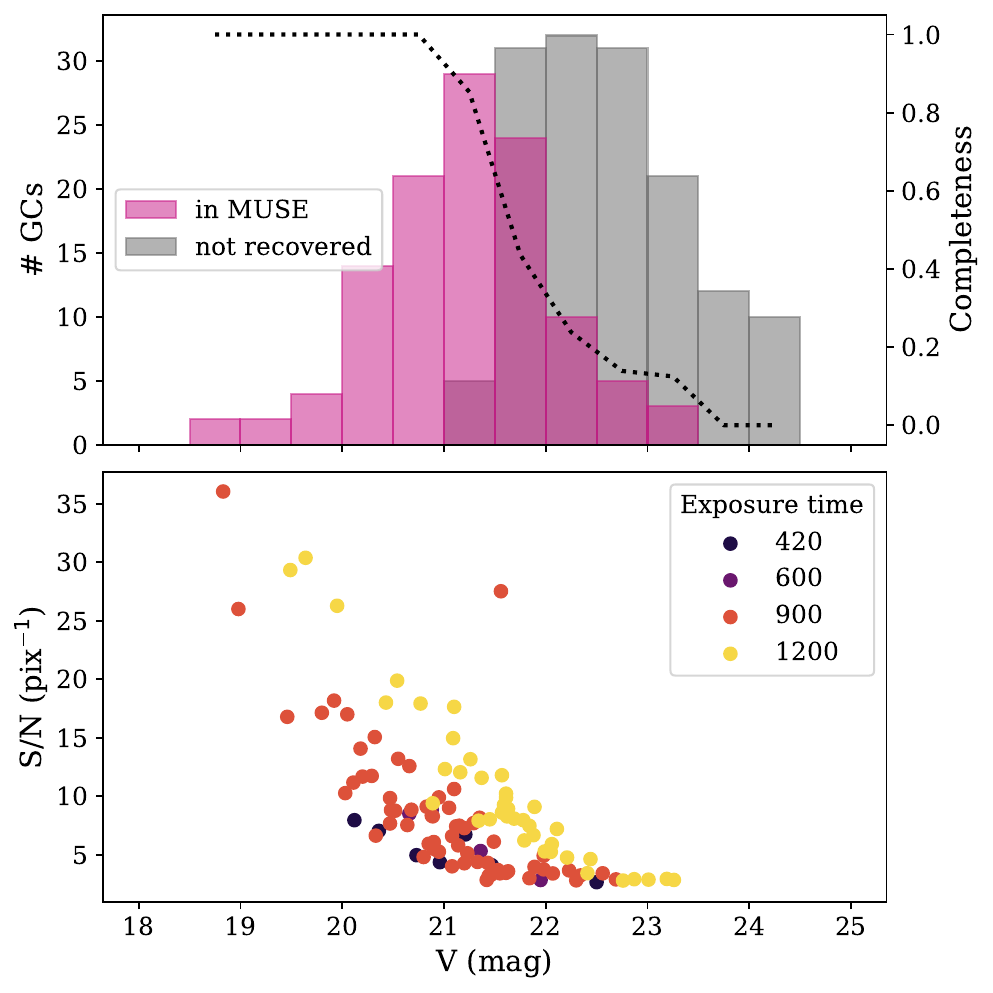}
    \caption{Completeness and signal-to-noise of the GCs recovered in the MUSE data. Top: Histogram showing the $V$-band magnitudes of the GCs in MUSE (pink) that have been cross-matched with the HST ACS catalogue \citep{Spitler2006, Harris2010}. GCs listed in the HST catalogue that were not recovered in MUSE are shown in the grey histogram. The black dashed line shows the fraction of recovered over all HST ACS GCs as a measure of completeness (right y-axis). Bottom: Signal-to-noise ratio of the MUSE GCs as a function of $V$-band magnitude. The total exposure time of the corresponding MUSE pointing is shown by the colour. }
    \label{fig:MUSE_GCs_completeness}
\end{figure}

To check how well our approach performed in extracting GC spectra from MUSE data, we compared the $V$-band magnitudes of MUSE GCs with HST match with all GCs located in the MUSE field of view from the HST catalogues in the top panel of Fig. \ref{fig:MUSE_GCs_completeness}. As the two histograms show, we can recover all of the GCs brighter than $V < 21$ mag from the MUSE data. As a measure of completeness, we show also the fraction of GCs in MUSE over all GCs listed in the HST catalogue (a measure of the completeness). Based on this, we find a completeness of 50\% at $V \approx 21.5$ mag. 

In the bottom panel of Fig. \ref{fig:MUSE_GCs_completeness}, we show the $S/N$ of the MUSE GCs as a function of their $V$-band magnitude. The points are colour-coded by the total exposure time of the MUSE pointing position in which the GCs are located, showing that the GCs in the fields with larger exposure times consistently have higher $S/N$ values, as expected. However, there is considerable scatter, likely due to the varying galaxy background brightness. Consequently, also the completeness shown in the top panel is a complex function of intrinsic GC brightness, exposure time, and galaxy background brightness. 

\section{Velocity and metallicity distributions in the central region}
\label{app:central_distribution}
\begin{figure}
    \centering
    \includegraphics[width=0.45\textwidth]{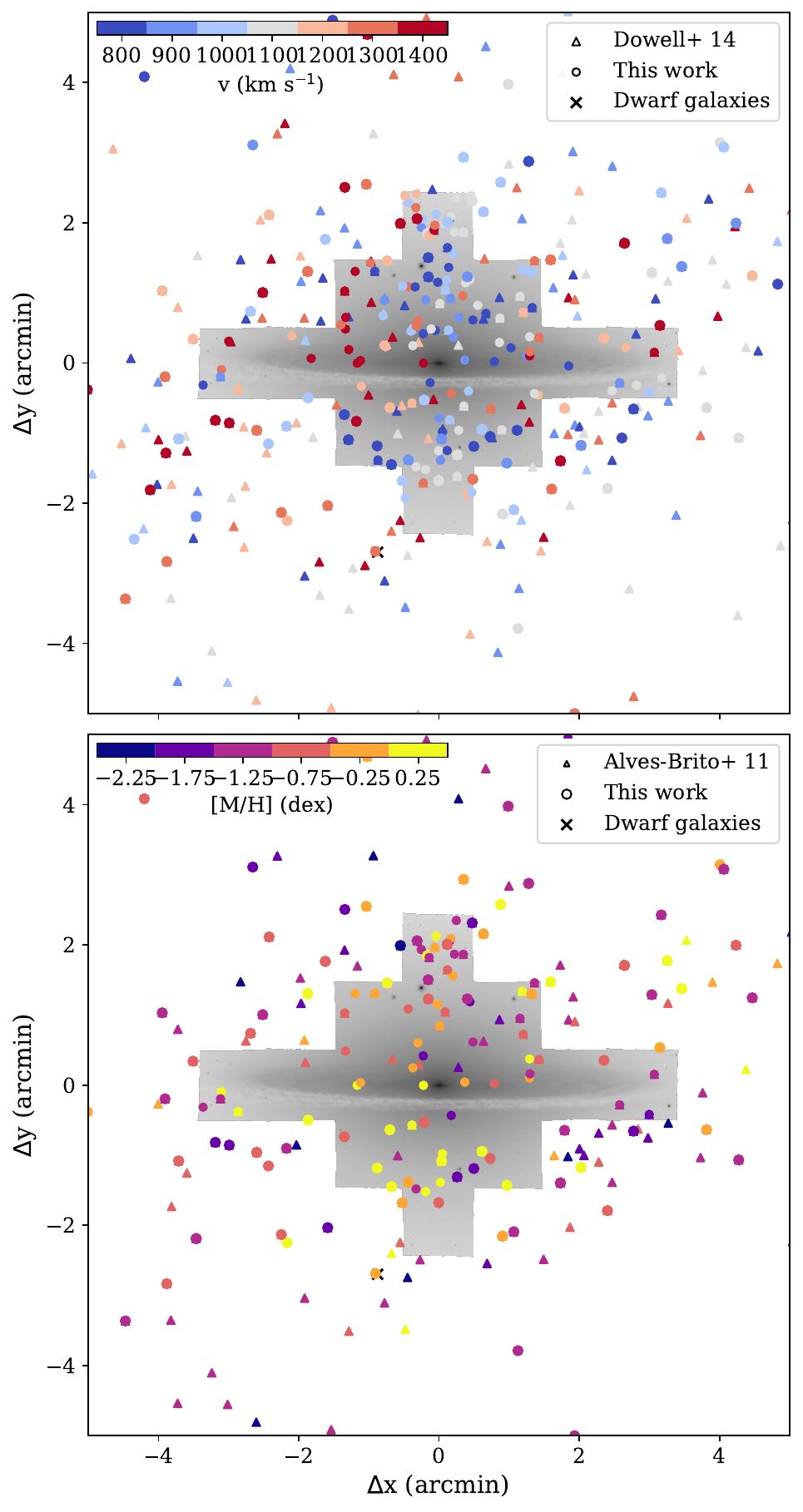}
    \caption{Spatial distribution of GC velocities (left) and metallicities (right), in the central 5\arcmin $\times$ 5\arcmin. The MUSE white light image is shown in grey.  
     Circles refer to this work, triangles refer to GCs listed in the \citealt{Dowell2014} and \citealt{AlvesBrito2011}. The black cross marks the position of SUCD1 \citep{Hau2009}.}
    \label{fig:GC_vel_metal_sky_alt}
\end{figure}

As an addition to Fig. \ref{fig:GC_vel_metal_sky}, Fig. \ref{fig:GC_vel_metal_sky_alt} shows the spatial distributions of velocities and metallicities in the central 5\arcmin $\times$ 5\arcmin.

\section{Iron versus total metallicities}
\label{app:EMILES_vs_XSL}
\begin{figure}
    \centering
    \includegraphics[width=0.46\textwidth]{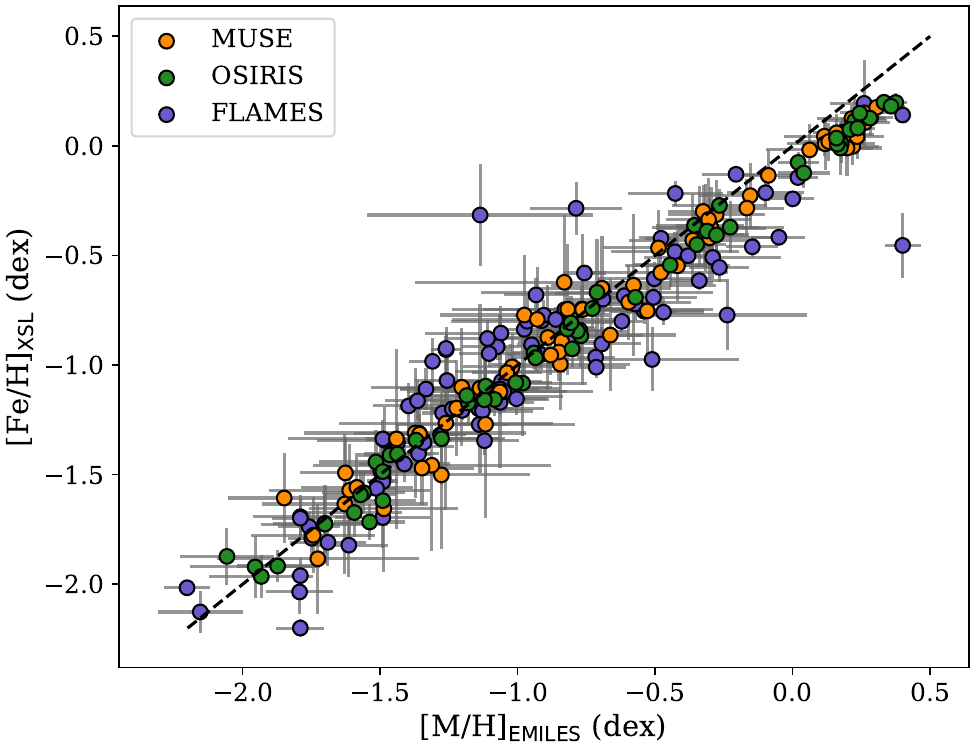}
    \caption{Total metallicities from fitting the spectra with E-MILES SSPs versus iron metallicities from the XSL SSPs. The different colours show GCs from OSIRIS, MUSE, and FLAMES. The dashed line is the one-to-one relation.}
    \label{fig:EMILES_vs_XSL}
\end{figure}

As described in Sect. \ref{sect:fitting_MUSE_spectra}, we used both the E-MILES and the XSL SSPs to fit the spectra from MUSE, OSIRIS, and FLAMES for testing purposes. In Fig. \ref{fig:EMILES_vs_XSL}, we show a comparison between the derived total metallicities from the E-MILES models and the iron metallicities from the XSL SSPs. As this figure shows, the two models agree very well, especially when considering the uncertainties. Only for high metallicities did we observe deviations, and those are likely caused by the differences in the given metallicity grids. 

\section{An updated catalogue of spectroscopically confirmed GCs around M\,104}
\label{app:catalogue}
Building on the extensive collection reported in \cite{Dowell2014}, we provide here an updated spectroscopic catalogue of GCs around M\,104 available combining literature GCs with our measurements from MUSE, FLAMES, and OSIRIS. Table \ref{tab:catalogue} shows an excerpt of the full catalogue that will be available in machine-readable form on CDS. For illustration purposes, we show here a handful of GCs that are at least in two of our three samples.

Here we briefly review the references that have contributed to this catalogue. As described before, we based our catalogue on the one presented in \cite{Dowell2014}. In their work, they presented spectroscopy for 152 unique objects based on spectra from the AAOmega spectrograph on the Anglo-Australian Telescope (AAT) and the Hydra spectrograph at the WIYN telescope at Kitt Peak National Observatory. The remaining sample stems from various sources including 46 velocity measurements from \cite{Bridges1997} based on spectra from the Low Dispersion Survey Spectrograph at the William Herschel Telescope, 16 from \cite{Larsen2002} using Keck Low Resolution Imaging Spectrometer observations, 170 from \cite{Bridges2007} based on data from the Two-Degree Field (2dF) spectrograph at the ATT and WIYN/Hydra data, and 259 from \cite{AlvesBrito2011} that are based on data from the DEep Imaging Multi-Object Spectrograph (DEIMOS) data on Keck. Removing duplicates, the final catalogue of \cite{Dowell2014} lists 579 unique objects. Based on their contamination analysis, they estimated that about 360 of those are genuine GCs of M\,104. 

In addition to velocities and associated uncertainties, the \cite{Dowell2014} catalogue also lists photometric measurements in $V$ as well as $B-V$ and $B-R$ colours from \cite{RhodeZepf2004} based on $B, V$, and $R$ imaging with Mosaic detector on the Kitt Peak 4m telescope. We adopt those photometric measurements, but combined them with the HST ACS $B$, $V$, and $R$ photometry described in \cite{Spitler2006} and \cite{Harris2010} where available. Additionally, we matched our catalogue to the photometric catalogue from \cite{Kang2022} that provides $u$, $g$, and $i$-band magnitudes obtained with MegaCam at the Canada-France-Hawaii Telescope.

\begin{sidewaystable}[]
    \centering
    \caption{Globular cluster catalogue. The full version is available at the CDS.}
    \begin{tabular}{c c c c c c c c c c c} \hline\hline

    ID & RA & DEC & $v_\text{MUSE}$ & $v_\text{OSIRIS}$ & $v_\text{FLAMES}$ & $v_\text{Dowell+}$ & [M/H]$_\text{MUSE}$ & [M/H]$_\text{OSIRIS}$ & [M/H]$_\text{FLAMES}$ & V \\ 
    & (J2000) & (J2000) & (km s$^{-1}$) & (km s$^{-1}$) & (km s$^{-1}$) & (km s$^{-1}$) & (dex) & (dex) & (dex) & (mag) \\ \hline
68 & 189.99769 & $-$11.65104 & 1119.1 $\pm$ 10.6 & 1157.2 $\pm$ 64.6 & ... & ... & $-$0.60 $\pm$ 0.23 & ... & ... & 21.01 \\
71 & 189.98116 & $-$11.64691 & 949.6 $\pm$ 14.1 & 981.8 $\pm$ 31.6 & ... & ... & ... & 0.23 $\pm$ 0.12 & ... & 20.91 \\
75 & 189.98917 & $-$11.64292 & 733.9 $\pm$ 12.6 & ... & 721.9 $\pm$ 0.4 & ... & $-$1.85 $\pm$ 0.20 & ... & $-$2.20 $\pm$ 0.08 & ... \\
113 & 189.95761 & $-$11.61719 & 1245.1 $\pm$ 7.1 & 1290.9 $\pm$ 17.4 & ... & ... & $-$0.84 $\pm$ 0.16 & $-$0.80 $\pm$ 0.09 & ... & 18.98 \\
258 & 190.00654 & $-$11.65105 & 947.6 $\pm$ 13.1 & 852.1 $\pm$ 25.8 & ... & 1045.0 $\pm$ 7.0 & $-$0.31 $\pm$ 0.16 & ... & ... & 20.89 \\
442 & 190.00504 & $-$11.64614 & 918.1 $\pm$ 16.8 & 932.2 $\pm$ 17.9 & ... & 897.0 $\pm$ 10.0 & $-$0.30 $\pm$ 0.26 & $-$0.28 $\pm$ 0.12 & ... & 20.95 \\
413 & 190.00021 & $-$11.60261 & 401.7 $\pm$ 3.9 & ... & 399.5 $\pm$ 0.4 & 392.0 $\pm$ 8.0 & $-$0.88 $\pm$ 0.10 & ... & $-$1.34 $\pm$ 0.07 & 18.83 \\
423 & 189.99700 & $-$11.64114 & 937.6 $\pm$ 6.4 & ... & 936.3 $\pm$ 0.9 & 947.8 $\pm$ 7.2 & 0.22 $\pm$ 0.08 & ... & $-$1.24 $\pm$ 0.08 & 20.29 \\
424 & 189.96816 & $-$11.64635 & ... & 1568.8 $\pm$ 16.3 & 1618.7 $\pm$ 1.1 & 1585.0 $\pm$ 8.0 & ... & $-$1.01 $\pm$ 0.13 & $-$1.51 $\pm$ 0.07 & 20.13 \\
414 & 190.06078 & $-$11.64111 & ... & 944.0 $\pm$ 13.2 & 996.3 $\pm$ 0.3 & 999.0 $\pm$ 8.0 & ... & $-$0.57 $\pm$ 0.10 & $-$0.61 $\pm$ 0.08 & 18.83 \\
276 & 189.99562 & $-$11.58962 & 1045.1 $\pm$ 6.9 & 1029.5 $\pm$ 10.7 & 1038.3 $\pm$ 0.6 & 1034.3 $\pm$ 7.6 & $-$0.85 $\pm$ 0.13 & $-$0.93 $\pm$ 0.09 & $-$0.62 $\pm$ 0.08 & 19.49 \\\hline
    \end{tabular}
    
    \label{tab:catalogue}
\end{sidewaystable}

\end{document}